\documentclass[manuscript]{acmart}

\AtBeginDocument{%
  }

\setcopyright{acmlicensed}
\copyrightyear{2018}
\acmYear{2018}
\acmDOI{XXXXXXX.XXXXXXX}

\acmConference[**'**]{Make sure to enter the correct
  conference title from your rights confirmation emai}{June 03--05,
  2025}{Woodstock, NY}
\acmISBN{978-1-4503-XXXX-X/18/06}

\usepackage{subcaption}

\begin{document}

\title[LLM-Powered Question Generation]{``Real Learner Data Matters'' Exploring the Design of LLM-Powered Question Generation for Deaf and Hard of Hearing Learners}

\author{Si Chen $^*$}
\email{sic3@illiois.edu}
\orcid{0000-0002-0640-6883}
\affiliation{%
  \institution{School of Information Sciences, University of Illinois Urbana-Champaign}
  \city{Champaign}
  \state{Illinois}
  \country{USA}
  \postcode{61802}
}

\author{Shuxu Huffman $^*$}
\email{shuxu@deaf.codes}
\affiliation{
  \institution{Gallaudet University}
  \city{Washington}
  \state{District of Columbia}
  \country{USA}
}

\author{Qingxiaoyang Zhu $^*$}
\email{qinzhu@ucdavis.edu}
\affiliation{
  \institution{University of California, Davis}
  \city{Davis}
  \state{California}
  \country{USA}
}

\author{Haotian Su}
\email{haotias@g.clemson.edu}
\affiliation{
  \institution{Clemson University}
  \city{Clemson}
  \state{South Carolina}
  \country{USA}
}

\author{Raja Kushalnagar}
\email{raja.kushalnagar@gallaudet.edu}
\affiliation{
  \institution{Gallaudet University}
  \city{Washington}
  \state{District of Columbia}
  \country{USA}
}

\author{Qi Wang}
\email{qi.wang@gallaudet.edu}
\affiliation{
  \institution{Gallaudet University}
  \city{Washington}
  \state{District of Columbia}
  \country{USA}
}



\begin{abstract}
Deaf and Hard of Hearing (DHH) learners face unique challenges in learning environments, often due to a lack of tailored educational materials that address their specific needs. This study explores the potential of Large Language Models (LLMs) to generate personalized quiz questions to enhance DHH students’ video-based learning experiences. We developed a prototype leveraging LLMs to generate questions with emphasis on two unique strategies: Visual Questions, which identify video segments where visual information might be misrepresented, and Emotion Questions, which highlight moments where previous DHH learners experienced learning difficulty manifested in emotional responses. Through user studies with DHH undergraduates, we evaluated the effectiveness of these LLM-generated questions in supporting learning experience. Our findings indicate that while LLMs offer significant potential for personalized learning, challenges remain in the interaction accessibility for the diverse DHH community. The study highlights the importance of considering language diversity and culture in LLM-based educational technology design.
\end{abstract}


\maketitle
\def\thefootnote{*}\footnotetext{These authors contributed equally to this work}\def\thefootnote{\arabic{footnote}}

\section{Introduction}

Deaf and Hard of Hearing (DHH) students face unique learning challenges compared to their hearing peers. These challenges often stem from differences in learning styles, such as a stronger reliance on visual learning methods, as well as the linguistic diversity present within the DHH learners \cite{Drigas2005elearning,Reed2008academicstatus,chen2024signmaku}. The variability in language backgrounds—ranging from American Sign Language (ASL) to various levels of written language proficiency \cite{Bat-Chava2000diversity, Scott2021callfordiversity,chen2024mixed}—further complicates the learning experience for DHH students, requiring specialized approaches to meet their learning needs . 

Creating effective learning materials for DHH students, and for special education in general, demands a significant amount of expertise and experience. The process is often highly personalized to the specific needs of each student \cite{alnahdi2024enhancing,holman2024navigating}, making the creation of these materials time-consuming and sometimes inconsistent \cite{scardamalia2019consistently,mooney2015love}. 
The challenge lies not only in the effort required to create personalized content but also in ensuring that it is free from bias, especially with the increasing use of AI in content generation. For example, research shows there are significant biases in the current state of sign language recognition, generation, and translation AI research such as an overfocus on addressing perceived communication barriers, and a lack of use of representative datasets \cite{desai2024systemic}. Mack et al. found that generative AI models consistently produced reductive archetypes for various disabilities, reinforcing broader societal stereotypes and biases \cite{mack2024wheelchair}.

With the advancement of Large Language Models (LLMs), their applications are increasingly being used in education to personalize learning and provide tutoring. LLMs offer remarkable flexibility in creating contextually appropriate material and excel in language fluency \cite{kasneci2023chatgpt, team2023gemini, touvron2023llama}. Their ability to generate tailored questions, answers, and other learning content to enhance learner experience is being explored by researchers and practitioners in hearing dominant education settings \cite{lu2023readingquizmaker}. 
However, whether such LLMs can generate personalized learning content and questions tailored to the needs of DHH students with diverse language capabilities and communication preferences remains to be studied. By leveraging LLMs, educators could potentially develop more accessible and personalized learning experiences, bridging some of the gaps that currently exist in special education.

Our study explores LLM-powered quiz question generation catered to DHH undergraduate students in video-based learning. Video-based learning, frequently available via YouTube or MOOC, is an educational approach that uses videos as the primary medium to deliver content and facilitate learning\cite{Khasawneh2023videomedia,Debevc2004roleofvideo}. It involves the use of recorded or live video presentations, tutorials, lectures, and demonstrations to teach concepts, skills, or convey information. This approach leverages the visual and auditory elements of videos to enhance understanding and engagement, making it a versatile tool for various learning environments. When it comes to studying DHH students' video-based learning, caption accuracy and design are often mentioned, e.g.,\cite{bhavya2022collaborativecaption,lasecki2014caption,kushalnagar2012readability}. However, this particular research explores question generation and studies the possibilities of personalized learning experiences associated. 

Our study answers two research questions. 
The first research question focuses on understanding users' perceived benefits and suggestions towards our prototype that applies LLM to generate different types of quiz questions for DHH students. The second research question uses the prototype as a starting point to understand valuable feedback in LLM-powered interaction for DHH learners. The second research question is highly relevant to the viability and acceptance of the prototype studied in the first research question.

\textbf{RQ1: How do DHH learners perceive our prototype for learning?}

\textbf{RQ2: What are the challenges faced by DHH learners in LLM-powered interactions?} 

Our work makes the following contributions: (1) design, implement, and evaluate a prototype that leverages LLM to generate questions catered to DHH learners; (2) propose two new types of questions helpful for DHH learners based on their unique learning needs: \textit{Visual Questions} (identify video timestamps where visual information is more likely to be missed) and \textit{Emotion Questions} (identify video timestamps where DHH learners have shared emotions reactions to the video content collected in previous research 
); and (3) understand whether LLM-powered learning experience are truly beneficial and accepted by the DHH community, rooted in non-technical and broader social issues such as language diversity. 
This includes considerations in LLM access, user English language proficiency, potential technology bias towards users whose first language is not English, and inclusive design that makes users fully embrace the learning technology.

\section{Related Works}

\subsection{DHH Learning with Video-Based Materials}
Online video-based learning presents additional challenges and difficulties for DHH learners, which supports the need for higher education institutions to enhance distance learning platforms to ensure equitable access and success \cite{alshawabkeh2024covid}. Researchers have investigated the factors that facilitate or hinder the use of video media in distance learning for deaf students and have explored the solutions that have been implemented to address these issues \cite{Khasawneh2023videomedia}. They have also examined the effectiveness of web-based video lectures for deaf learners compared to traditional teaching methods that use a sign language interpreter, particularly in the contexts of education and rehabilitation \cite{Debevc2004roleofvideo}.

Similar studies have been conducted internationally. For example, researchers in Indonesia developed digital video-based learning media for rampak kendang instruction, which utilizes sign language to support deaf learners \cite{Pratiwi_2019}. Other researchers have designed a learning system that provides Greek Sign Language videos alongside text for vocational and educational training \cite{Drigas2005elearning}.

The Deaf community is diverse, with members having varied backgrounds, identities, cultures, and language abilities \cite{Bat-Chava2000diversity, Scott2021callfordiversity}. Given this diversity, it is crucial for educators to consider individual and school-level variables, evaluate outcomes on comparable groups, and utilize diverse instructional methodologies to enhance student learning and knowledge acquisition based on his/her learning style and preference. Improve research in bilingual and multilingual deaf education to accommodate individual needs is critical. 

\subsection{DHH Learning with Emerging Technology}

Recent studies have focused on utilizing various emerging technologies and tools to enhance the educational experiences of DHH students. Previous studies have evaluated the academic status of DHH students in public schools and identified factors that help or hinder their academic success \cite{Reed2008academicstatus}. Prior research has also highlighted the need to adapt educational technologies to address the unique needs of DHH learners \cite{alshawabkeh2024covid, wang2018a11y}. 

Researchers have explored various solutions involving emerging technologies for DHH students' learning experiences, such as eye tracking, to better understand gaze patterns, attention management, and communication strategies among DHH individuals \cite{agrawal2021eyetracking, sedlackova2020textbook}. These studies have informed textbook development and visual split attention management strategies. Additionally, innovative solutions like instructional mobile games have been investigated, demonstrating positive attitudes and skill transfer among DHH learners, along with benefits such as enhanced orienteering skills and peer tutoring \cite{shelton2016math}.

Various technologies in the classroom have been studied extensively. One example is Virtual Reality (VR), which has been utilized to create immersive classroom experiences for DHH individuals. These experiences incorporate live and recorded sign language interpretation to improve access and collaboration in MOOCs \cite{paudyal2018daveeVR}. For American Sign Language (ASL) users, signing avatars have been employed using VR to support students with hearing loss, particularly those in rural schools with limited access to specialized educators \cite{zirzow2015avatar}. Researchers have also developed "Signmaku," an ASL-based commenting feature that enriches the co-learning experience for DHH individuals \cite{chen2024signmaku}.

Beyond VR, other technologies have also been explored. Examples include Internet of Things (IoT) and Augmented Reality (AR) technologies, which have been investigated for DHH learners in e-learning platforms. Researchers explored how integrating IoT and AR can enhance learning for the deaf by analyzing data, interfaces, and pervasiveness, and proposed a deaf intelligent learning model based on a preliminary survey of deaf learners \cite{alrashidi2023synergistic}. Additionally, research targeting ASL users in virtual think-aloud sessions investigates the behaviors of DHH participants to understand the challenges of using the think-aloud method and identify issues such as asynchrony between signing and navigating interfaces and the use of ambiguous visual descriptive signs. This provides methodological and design implications for more effective virtual think-aloud studies with DHH participants \cite{chen2023thinkaloud}.

For audio-based classrooms, researchers have examined how real-time captions can support DHH students \cite{lasecki2014caption, kushalnagar2012readability}. The use of Automatic Speech Recognition (ASR) has been explored, revealing both its benefits and challenges in DHH classrooms \cite{kushalnagar2012readability, butler2019speechrecognition}. Group autoethnography has been conducted in mixed-hearing ability, higher education settings to explore the effectiveness of technologies like live ASR and typing strategies to promote inclusive collaboration as well as design implications for accessible communication technologies \cite{chen2024mixed}. Researchers have also investigated how learners edit captions of educational videos and highlighted the benefits of crowd-sourcing for accuracy and cost-efficiency \cite{bhavya2022collaborativecaption}.

In this fast-moving world, particularly with the rise of AI, the deaf community faces both unprecedented opportunities and risks, given the unique language and cultural aspects of this group. For instance, AI development, which relies heavily on large datasets, raises significant Fairness, Accountability, Transparency, and Ethics considerations, particularly concerning the personal nature of sign language data and its cultural significance to the deaf community \cite{bragg2021fate}. Moreover, sign language AI studies can introduce unconscious bias, often driven by hearing researchers, focusing on perceived communication barriers, using non-representative datasets, and building on flawed models \cite{desai2024systemic}. The question of AI fairness for people with disabilities, particularly DHH individuals, extends beyond inclusive data to encompass considerations of interpretability, ethical responsibilities, and appropriate evaluation metrics, as well as understanding the impact of AI on human behavior and user abilities \cite{Kafle2019fairness}. Additionally, there remains a significant opportunity to explore AI in education for the deaf community, as curiosity and concerns about the surge of AI persist within this group \cite{Coy2024education}.

Our research seeks to explore the potential of AI to provide more personalized learning experiences for DHH learners by addressing the diverse language and learning preferences within this community. By focusing on adaptability and customization, we aim to contribute to the development of educational technologies that better align with the varied needs of DHH learners, enhancing accessibility and engagement.

\subsection{Automatic Language Generation for Learning Materials}

Learning materials, such as textbooks, videos, and practice exercises, play a crucial role in supporting learners' experiences and meeting course objectives. However, constructing learning materials at scale that cater to the diverse needs of individual learners remains a significant challenge. Previous research has explored the use of automatic generation methods to create personalized learning content or to adapt original materials into more individualized formats~\cite{draxler2023relevance, chen2024gptutor, choi2024vivid}. Efforts have also been made to develop practice exercises suited to varying proficiency levels~\cite{cui2023adaptive, stowe2022controlled}, and to provide real-time contextual feedback for learners during the learning process~\cite{nagata2021shared, nagata2019toward, fung2024automatic}. However, potential gaps exist, such as the alignment between users' expectations of generated materials and the effort required to integrate these materials into real-world learning environments. To better understand the challenges and opportunities of creating personalized learning experiences using automatically generated materials, we focus on a specific learning material -- in-video prompting questions.

In-video prompting questions is a critical means to improve learners’ learning performance in video learning~\cite{shin2018understanding, van2021effects}. Such learning prompts effectively help learners maintain their engagement and review the video content while watching the video~\cite{karpicke2011retrieval}. Previous work has also shown the effectiveness of prompting questions on facilitating DHH learners' video learning process by increasing their active engagement with the materials~\cite{banda2011review, callender2007benefits}. However, generating such questions is time- and resource-consuming for instructors who need to spend a huge amount of time identifying where to generate a question, what and how to ask in the question, making the process of question generation arduous~\cite{niebuhr2014online}. Due to the heavy drain on instructors for their time and resources, the prompt questions are therefore less accessible to students at scale. 

Existing work has explored generating in-video prompting questions through instructional design knowledge, crowd-sourcing methods, and automatic NLP algorithms~\cite{lu2023readingquizmaker, wang2019upgrade, yang2024aqua, lindberg2013generating}. Promisingly, the automatic question-answering generation has the potential to save time and human effort, making prompting questions accessible to students at scale. Before the prevalence of large language models (LLMs), questions were generated by rule-based systems or non-pre-trained neural networks based on key phrases extracted~\cite{kurdi2020systematic, lindberg2013generating}, resulting in low-quality and diversity issues. While LLMs were available for generating diverse text sequences, for example in various lengths, using a wide range of vocabularies, and in different styles, more trials have been conducted to leverage LLMs for question generation. Some researchers proposed the criteria and evaluation metrics for LLMs to generate questions for learning mastery~\cite{hutt2024scaling}, while others focused on generating and assessing multiple-choice questions based on video transcripts~\cite{arif2024generation}. However, less attention has been paid to the specific needs of student groups, such as DHH learners which are of particular interest to us. 

In this work,  we aim to explore how to leverage the automatic question generation technique for DHH learners‘ needs, what the limitations of the question generation process are, and how to overcome these limitations. We leverage LLMs as a means for automatic question-answering generation for two reasons: LLMs can accommodate diverse language ability, which allows DHH learners to customize the question language based on their proficiency and the interactive generation process allows iterative question generation and adaptation based on users’ needs. 
\section{Methods}


To investigate and demonstrate how the co-generation pipeline works, the research team leveraged a free mainstream learning video from Coursera as the input data to generate questions. The video, marked as beginner-level and with a positive review score (average score of 4.5 out of 5 by 3.7k learners), delivers instructions on the topics of augmented reality (AR). The team selected the first lesson (less than 15 minutes) in the video that covered the history of AR. Since the video is designed for hearing students, the instructor uses spoken English to teach. The video presentation style alternates 15 times between a talking head and full-screen visual examples and sometimes includes both simultaneously. The research team added error-free captions to the video.

To better understand the requirements of the target learners' needs before customizing prompts for the LLM to generate questions for DHH learners who would watch the learning video, researchers conducted a preliminary discussion with two professors who teach and conduct research with DHH learners and two DHH undergraduate students. The discussion focused on four aspects of the conventional question generation process: where to generate the questions, what to ask in the questions, how to ask the questions, and when to ask the questions. The discussion resulted in a set of focal points including the purpose, type, language level, and difficulty of questions and when to pop up the questions. Based on the responses from the experts and DHH students, researchers summarize the major criteria for the LLM prompts to generate questions tailored to DHH learners' needs. They include: generate questions that (1) are answerable based on the video content; (2) reflect important concepts and contents or difficult content that often confuse students (in this case, important and/or difficult content is echoed via question prompting, and students can be nudged to pay attention to and rethink those content~\cite{cotton1988classroom}); (3) are closed questions, such as True-or-False (T/F) and Multiple Choice (the professors who are experienced teaching and working with DHH learners suggested using closed questions as a starting point so that DHH learners could focus on the video learning content and minimize their effort in the typing/writing process); and (4) simple language syntax (sentence/vocabulary) to reduce unnecessary confusion and cognitive overload and maximize information clarity for better comprehension ~\cite{guardino2016deafness} (in this case, complex clauses, compound sentences, and double negatives should be minimized to ensure DHH learners with diverse language background can access LLM generated questions with ease). 
 
 

\begin{figure}[htp]

    \begin{subfigure}[b]{0.93\textwidth}
         \centering
         \includegraphics[width=\textwidth]{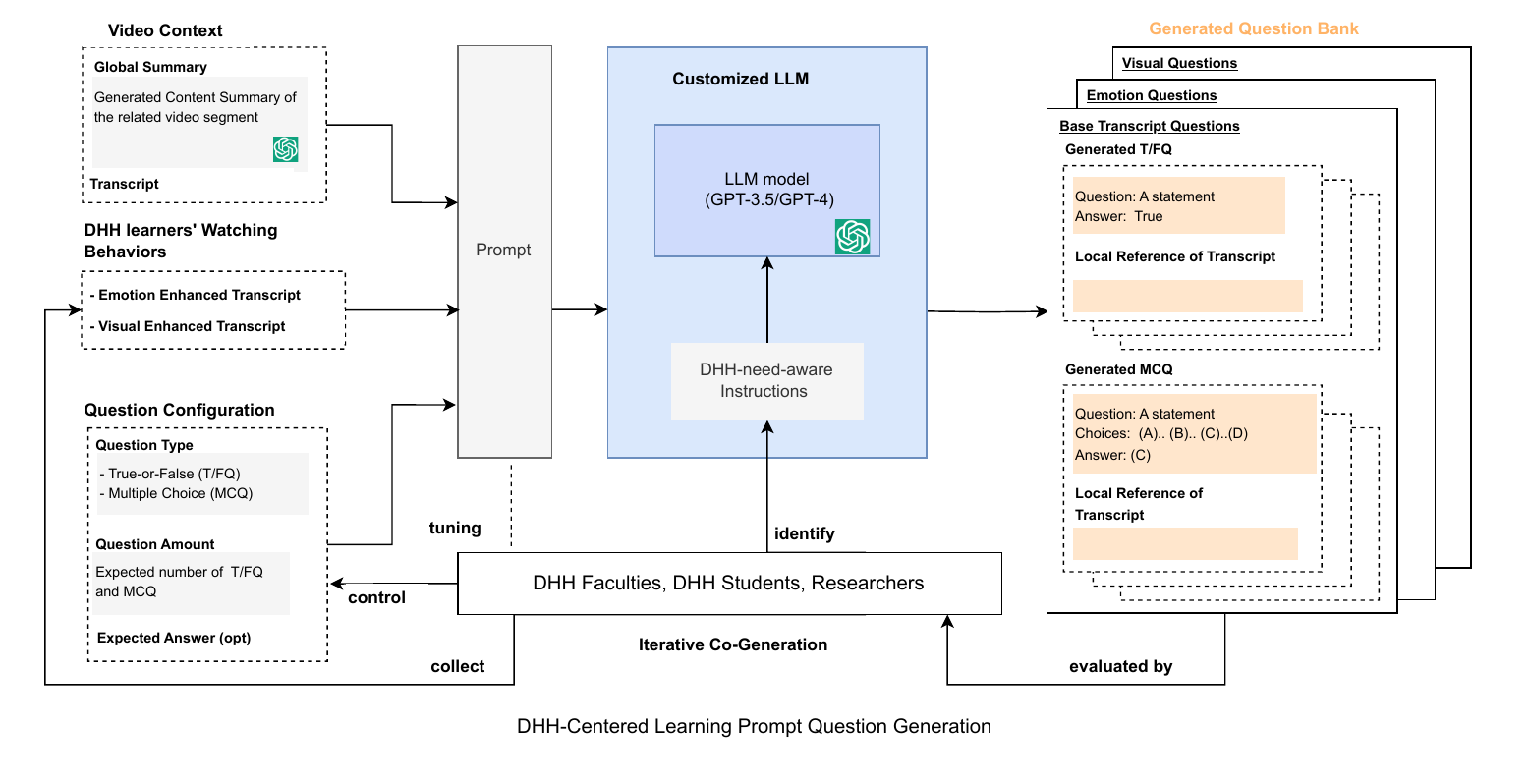}
     \end{subfigure}
     \hfill
     \begin{subfigure}[b]{0.55\textwidth}
         \centering
         \includegraphics[width=\textwidth]{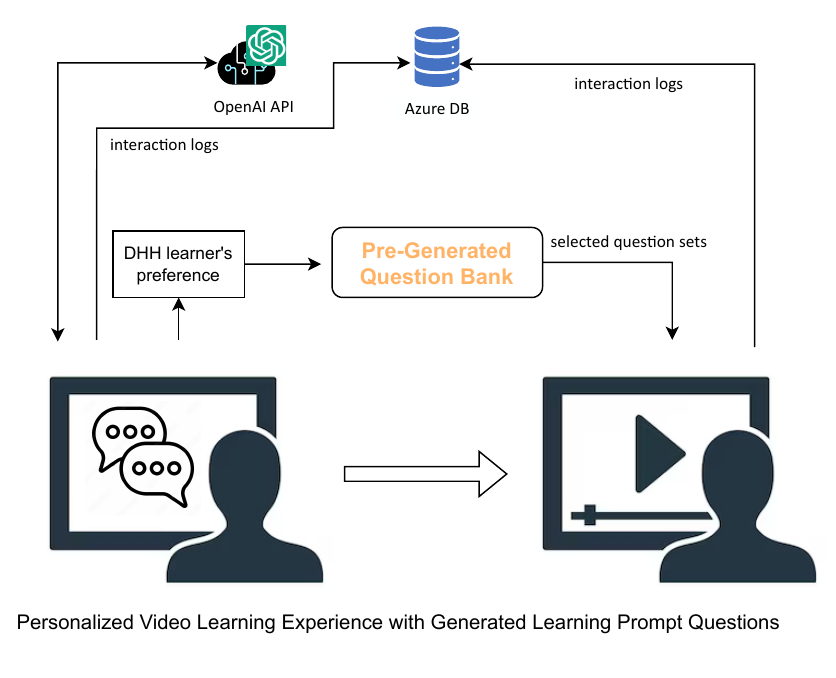}
     \end{subfigure}
     \hfill



\caption{
    The research team uses ChatGPT API to process video transcripts to generate quiz questions using three different strategies. The first strategy involves using only the transcript. The second adds emotional data at specific timestamps, and the third includes visual information from the video at corresponding points. Each strategy generates 10 candidate questions stored in a question bank, and students can choose one to three strategies to create a set of 10 questions they will answer in user study.} 
    \Description{}
\label{fig:question_gen}

\end{figure}


Based on these question generation guidelines/criteria, researchers could adjust LLM prompts and iteratively refine the LLM-generated questions until they align with the established guidelines. Additionally, researchers have initialized a 3-stage interactive question co-generation pipeline, allowing novice instructors to work with LLM to construct initial questions and iteratively optimize them with DHH learners’ needs in focus (as shown in Fig. \ref{fig:question_gen}):

\textbf{Stage 1 - Baseline Question Generation} 
\begin{itemize}

\item \textbf{[Strategy One]:  Transcript-only questions } At the first stage, the research team examined the potential and limitations of using LLM to generate learning questions for DHH learners. The researchers formalized a set of instructions for automatic question generation and prompt the LLM to follow the instructions to generate a set of learning questions. The generation process is solely based on the transcript and the given instructions. The LLM independently decides on the snippets it considers important and generates questions accordingly (\textbf{Transcript Questions}). Based on the generated initial questions, the team observed the outstanding text-processing ability of LLM~\footnote{Considering the quality and wide accessibility, the team used GPT-3.5 in the automatic question generation pipeline. Researchers also tested on diverse LLM backbones, such as GPT-4 and Gemini, the quality of GPT-3.5 is comparable to other LLMs for our task. Considering the on-par quality and higher accessibility of GPT-3.5, the team has utilized GPT-3.5 model. 
}. 
The LLM model is able to understanding the video transcript, extracting reference snippets of important content from the transcript and generate reasonable questions of required question types (e.g. the model is proficient at generating diverse and appropriate choices for multiple-choice question which is a tedious work for instructors). However, the LLM model is much less sensitive to the specific needs and characteristics of DHH learners as the datasets used to train the model lacks sufficient DHH related data. So the research team decided to include additional parameters by leveraging DHH learners' emotion enhanced transcript and visual enhanced transcript to generate questions. 
\end{itemize}
\textbf{Proposed Strategies for Generating Questions Incorporating both Transcript and DHH Learner Characteristics/Perspectives:}
The initial questions generated by the model without the input/perspective from DHH Learners is much less usable by the target users. To mitigate this deficiency, the research team has adopted the strategy to incorporate DHH learners’ perception and behavior data into the generation pipeline and include DHH learners and their instructors to evaluate the question drafts at the second stage. 
\begin{itemize}

\item \textbf{[Strategy Two]: Generation with Emotion-Enhanced Transcript} DHH learners’ behavior data is collected in previous work, their facial expression was captured while watching the learning video. Researchers aggregated time of periods with negative facial expressions from at least 3 participants, indicating the parts of content that confused them.  These snippets were injected into the generation pipeline to make the model aware of learners’ emotion during the learning. A new set of questions were generated based on the emotion data (\textbf{Emotion Questions}). 

\item \textbf{[Strategy Three]: Generation with Visual-Enhanced Transcript} Apart from behavior data, DHH learners’ perception data is also leveraged, which is collected in a previous work. In this dataset, 6 DHH learners and 2 instructors annotate the video frames (using time stamps) with hard-to-follow segments due to movements and captions (intensive visual movements, out of unalignment captions and so on). Researchers injected these hard-to-follow snippets into the generation pipeline to make the model aware of DHH learners’ visual perception of the learning video. Based on the prior knowledge of DHH learners’ visual perception, a new set of questions were constructed (\textbf{Visual Questions}). 
\end{itemize}

\textbf{Stage 2 - Baseline Question Evaluation: }
In order to further improve the question quality by seeking input from DHH learners and their instructors in the generation pipeline, the research team asked two DHH students (one identified as Deaf and the other as Hard of Hearing (HoH)) and one faculty closely working with DHH learners to evaluate the three sets of generated questions. They evaluated the question sets based on the following factors: question relevance, readability, timing appropriateness, and usability in earning and critical thinking enhancement. 

\textbf{Stage 3 - Baseline Question Revision:} Feedback from DHH learners and instructors in the second stage indicated the need to further improve question clarity by paraphrasing sentences, replacing words or choices to ensure their readability for most DHH learners with diverse English language capability and backgrounds.
As such, we iteratively improved the generated questions by prompting the LLM with instructions for simplification with detailed hints (e.g. use “attempt” instead of “foray”). The baseline questions were therefore shortened, simplified, manually validated, and saved in a test bank to be prototyped in the learning video for user studies.



\subsection{Prototype Design}
\begin{figure}[ht]
    \centering  
    \includegraphics[width=1.0\textwidth]{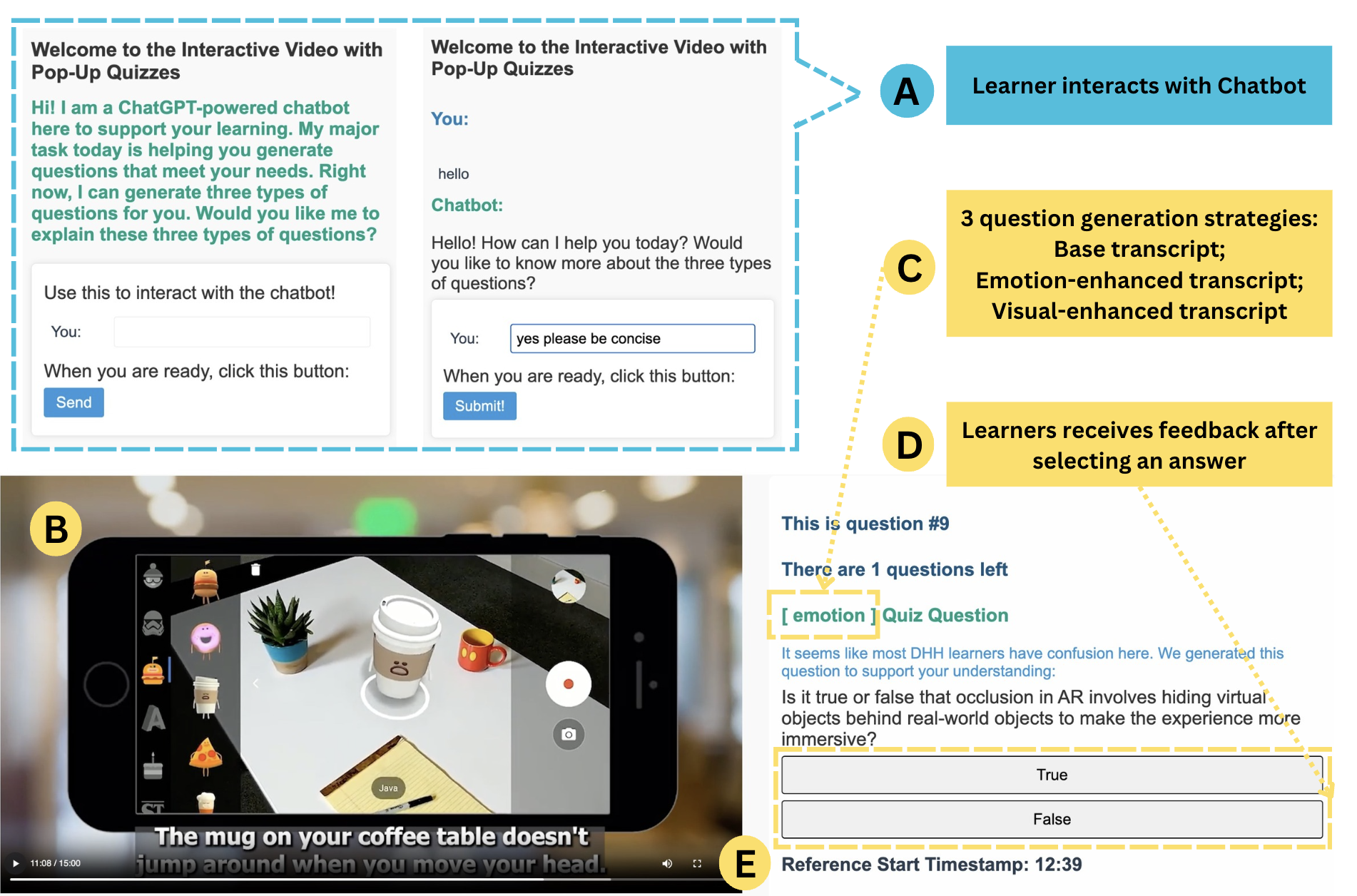} 
    \caption{Above, the chatbot greets participants by explaining the main task, which includes generating quiz questions using three different strategies, and prompts them to ask for clarification. Participants are encouraged to engage in conversation with the chatbot. When a participant types "hello," the chatbot redirects the conversation to focus on the questions generated with three strategies (A). Below and to the left, after the participant confirms 'emotion-enhanced transcript,', the video player displays the instructional video (B). To the right, the current question index number, the number of remaining questions for the video, the strategy on which the current question is based (C), the question itself (D), and the reference start timestamp (E) are indicated. There are three possible question-generating strategies based on the user’s interaction with the chatbot (A): Base transcript, emotion-enhanced transcript, and visual-enhanced transcript.} 
    \Description{}
    \label{fig:system_screens}
\end{figure}

Researchers developed and deployed a full-stack web application to facilitate participants' personalized learning experience with LLM assistance. To start with, participants interacted with OpenAI's ChatGPT\footnote{https://platform.openai.com/docs/overview} (Fig. \ref{fig:system_screens} A). The chatbot assisted participants in understanding the three question generation strategies deployed and answered their questions if any. Upon confirmation of their choices, the web application retrieved questions from the test bank (refer to the previous section) and registered the selected questions along with corresponding timestamps based on the participants' responses. It is important to note that, for this research, the questions are pre-generated by the LLM and hard-coded into the prototype (Fig. \ref{fig:question_gen}). In future iterations, we aim to dynamically generate questions in real-time. During video playback, the questions appear at the designated timestamps (Fig. \ref{fig:system_screens} B). After participants selected their answers, the web application would provide instant feedback (Fig. \ref{fig:system_screens} D). For correct answers, participants would receive encouraging remarks on the screen; whereas for wrong answers, participants were reminded to refer to the reference start timestamp and to rewind the video to find clues for answering the questions. When the learning session ended, participants were prompted to complete a onscreen survey to evaluate the questions they encountered. 

\subsubsection{Interactive LLM Chatbot for Question Personalization}

Researchers utilized OpenAI's GPT-4\footnote{https://openai.com/index/gpt-4/} to develop our chatbot, leveraging its capability to process contextual information and engage with participants effectively. The chatbot is integrated with an intuitive user interface, guiding participants through the interaction process. When participants access the web link, they are greeted with a simple guide embedded in a prompt template (Fig. \ref{fig:system_screens} A). The chatbot further interacts with participants to clarify the question types as needed. Participants can select one, two, or all three types (emotion, visual, and transcript), and the chatbot confirms their choices. Upon confirmation, the chatbot follows the prompt template to generate an output with specific tokens, such as \texttt{QUESTIONS transcript emotion visual DONE}, which match the predefined regular expression. The web application then selects questions from the pool based on the participants' choices, registers the questions, and proceeds to video display.

\subsubsection{Video Display with Question Pop-Up and Feedback}

Participants are then guided to watch the video (Fig. \ref{fig:system_screens} B). When a registered timestamp is reached, a pop-up question appears. The pop-up includes the question, the question number index, the remaining questions for the video, and the category of the selected question (Fig. \ref{fig:system_screens} C). Additionally, it shows the reference start timestamp (Fig. \ref{fig:system_screens} E). If participants forget or miss the reference from the video, they have the option to rewind to the specific point and watch again. Upon selecting an answer, participants receive feedback. If the selected answer is incorrect, the feedback directs them to the \texttt{Reference Start Timestamp} to revisit the relevant part of the video.

\subsection{User Study (N=16)}

\begin{figure}[ht]
    \centering  
    \includegraphics[width=1.0\textwidth]{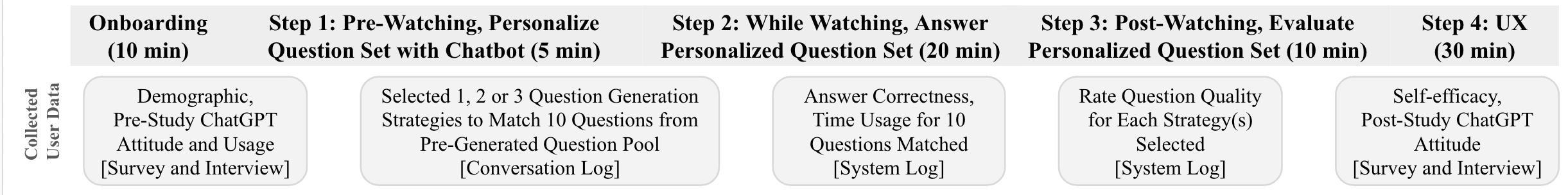}    
    \caption{User Study Process} 
    \Description{to add alt text}
    \label{fig:process}
\end{figure}

This is a mixed-method study to evaluate the prototype developed and implemented (refer to the previous sections) via user studies. Step 1 and Step 2 focus on user interactions with the prototypes; step 3 and step 4 center on user experience data collection. The study plan was approved by the university IRB. The process is visualized in Fig. \ref{fig:process}

The user study had 16 participants, including undergraduate and graduate students and deaf professionals.
Seven of the participants were in the 20-24 age group, two in the 25-30 age group, and six in the 31-42 age group. Nine identified as men, and seven identified as women. Ten identified as Deaf, and six identified as HoH. Six participants identified as White, one Middle Eastern, three Hispanic, five Asian, and three African-American. Nine noted ASL as their primary language and English as their second, while seven indicated English as their primary language and ASL as their second.
The research used both qualitative and quantitative analysis to explore differences in perceptions and experiences between Deaf and HoH learners. For many Deaf individuals, being deaf or hard of hearing is viewed as a cultural and linguistic distinction, which may impact their interaction with the prototype, as the interaction is presented in English. The participants were recruited through word of mouth and compensated at \$25 per hour for a total of \$50 for a 2-hour study. 

\subsubsection{Onboarding} Researchers explained the whole study process and collected participants' demographic information, current usage of, and opinions about ChatGPT. Researchers also collected participants' attitudes towards ChatGPT using survey questions by adapting the current ``Measurement of Negative Attitudes toward Robots'' scale \cite{nomura2006measurement}. A sample question looks like, "\textit{Would you feel uneasy if ChatGPT really had emotions?"} 

\subsubsection{Step 1: Chat with Chatbot to Personalize Question Set (Fig. \ref{fig:system_screens} - A)}
The first step is where the participant interacted with our chatbot powered by OpenAI’s API. The chatbot informed the participant about the three types of questions previously explained and asked the participant to select one of the three types he/she preferred to have for the video learning prototype. The participant was allowed to select one category, or any two, or all three. Once the choice was made, the prototype selected 10 questions from the test bank with appropriate types and would attempt to place questions such that their pop-up timestamps were distributed throughout the video, but otherwise random. Researchers collected data on user interactions with the chatbot, the questions set selected by participants based on their interaction with the chatbot.

\subsubsection{Step 2: Watch Video and Answer Pop-up Personalize Question Set (Fig. \ref{fig:system_screens}- B \& C \& D \& E)} The participant then watched a 15 minute video introducing the concept of augmented reality (AR), along with simple historical facts, current relevance, and basic attributes of AR. While watching the video, the prototype popped up the 10 questions selected during the onboarding process. If one question was not answered and where the video’s timestamp was equal to or past the question’s pop-up timestamp, then the prototype would pause the video and display the question on the side. The participant then must answer the question correctly to continue. The participant had the ability to rewind the video if he/she wished to. The participant could not fast forward past the next unanswered question. 
Researchers also collected action logs to better understand participants learning behavior- their answer attempts along with the time spent on each question, and correctness at first attempts xx and xx.

\subsubsection{Step 3: Post-Watching - Evaluate Personalize Quiz Question Set}\label{rate} Once the participants finished watching the video, they then evaluated the quiz questions that they read and answered while watching the video. These evaluation ratings take inspiration from multimedia learning theory on a 1-7 scale, and Bloom's taxonomy on a 1-7 scale intended to measure the effectiveness of the quiz questions in helping the participants understand the video content \cite{}. Mayer's multimedia learning theory \cite{mayer20143}articulates three types of cognitive demands as Essential Processing, Extraneous Processing, and Generative Processing. The three rating questions in our survey, \textit{Reducing Irrelevant Information}, 
\textit{Focusing on Essential Information},
\textit{Fostering Connection between Text and Image}, are adapted and adopted to the three demands based on the reading preferences of DHH learners in consultation with Deaf researchers and faculties. 

Sample rating questions include \textit{Overall, this group of questions helped me learn the video better by “Reducing Irrelevant Information” (1-7 Strongly Disagree to Strongly Agree)
}; \textit{Overall,  I think this group of questions helps me recall facts and basic concepts in this video (1-7 Strongly Disagree to Strongly Agree)
}. 

Participants were asked to rate \textit{each strategy} that generated quiz questions they selected, for example, if participants selected all three types, they would rate three times for the respective underlining strategies. Researchers conducted comparisons between rating outcomes on the three types of quiz questions in R2. But Step 3 in the study is to evaluate quiz question quality as part of user experience data.  Due to the various combinations and limited sample size, statistical tests were not feasible for comparing three-question generation strategies.

\subsubsection{Step 4: User Experience} To study participants' self-efficacy perceptions in using the experimental prototype, researchers first used a survey with eight questions on a 7-point Likert Scale (Strongly Disagree, Disagree, Slightly Disagree, Neither Agree or Disagree, Slightly Agree, Agree, Strongly Agree) in reference to previous works, e.g., \cite{bandura1997self}. Sample questions include, \textit{Using this prototype, I will be able to achieve most of the goals that I have set for myself better}. Additionally, a series of learning experience rating questions were designed to evaluate how the prototype enhanced the learning experience of the participants. Six 7-point Likert Scales, similar to the ones aforementioned, measured the extent to which the participants agreed with a series of statements covering various aspects of the learning experience. One example includes, \textit{Overall, This set of questions helped me learn the video better by focusing on essential information}. Descriptive analyses were performed on the two series of questions to obtain a fundamental understanding of how the prototype influenced the learning experience and enhanced the self-efficacy of the participants. To further understand how the learning experience with the prototype influenced self-efficacy, correlation analyses were performed between the question rating items and self-efficacy measurement. 

Using the same questions that appeared during onboarding, researchers again used a survey to confirm participants' attitudes towards ChatGPT in an attempt to learn how the experimental prototype might have changed users' thoughts and opinions on ChatGPT. Computed changes in attitude are shown in Fig. \ref{fig:attitude}. 

Finally, a semi-structured interview was conducted during which the researchers probed for feedback on the prototype. The researchers asked the participants about their comprehension of the video content, their thoughts about the quality of the pop-up quiz questions, and how they might like LLM to support personalized learning experiences. The research team conducted a thematic analysis of the qualitative results \cite{braun2012thematic}. Four researchers, two identified as Deaf, one as HoH, and one identified as hearing, independently reviewed 5-10 interview transcripts, noting common themes. The four researchers then collaboratively reviewed each other's notes and discussed the themes synchronously and asynchronously, due to language preferences and abilities, till a consensus was reached. Below present the findings of the agreed-upon themes, supported by participant quotes wherever needed. Common themes include attitude towards each step, comparison of three question generation strategies, suggestions for prototype improvement, and possibilities of using the prototype in real time if on-the-fly question generation was possible.

\subsection{Team Positionality} The research team includes members with diverse hearing and signing abilities: three identify as Deaf with fluent signing, one identifies as HoH with basic signing skills, two identify as hearing with some signing ability, and two identify as hearing with no signing ability. User studies were conducted by either Deaf members with fluent signing, HoH members with basic signing and oral English, or hearing members with some signing ability via human interpreters or in English, depending on participants' communication preferences. The two members with no signing ability participated in prototype design, development, and data analysis and did not take part in user studies. The prototype development and deployment were led by a Deaf researcher with fluent signing. The prototype design and data analysis were led by a hearing researcher with beginner signing ability. 

\section{Findings}

\subsection{RQ1-Perception of Prototype}

We used the quiz question-answering data collected in Step 2, the quiz question ratings collected in Step 3 and UX (survey and interview) collected in Step 4 to answer RQ1.  

\subsubsection{Improved Self-Efficacy and Learning Experience.}

Mapping the eight self-efficacy evaluation questions onto 7-point numerical scales from 1 to 7, the descriptive statistics showed that the tool generally improved the participants' self-efficacy ($N$ = 16, $M$ = 5.70), and all participants generally agreed that the prototype improved their self-efficacy (the minimal value of the average agreement level was 4.25 across participants, where `4' corresponds to a neutral attitude). In the question rating items, participants generally agreed that the questions enhanced learning by reducing irrelevant information, focusing on essential information, and building connection between text and image. Participants also generally concurred that the questions helped them recall facts and concepts, understand and explain concepts and ideas, and apply information to new situations. The analysis summaries are encapsulated in Fig. \ref{fig:value1} and Fig. \ref{fig:idrate}.

\subsubsection{Comparison of Three Question Generation Strategies}

\begin{figure}[ht]
    \centering  
    \includegraphics[width=1.0\textwidth]{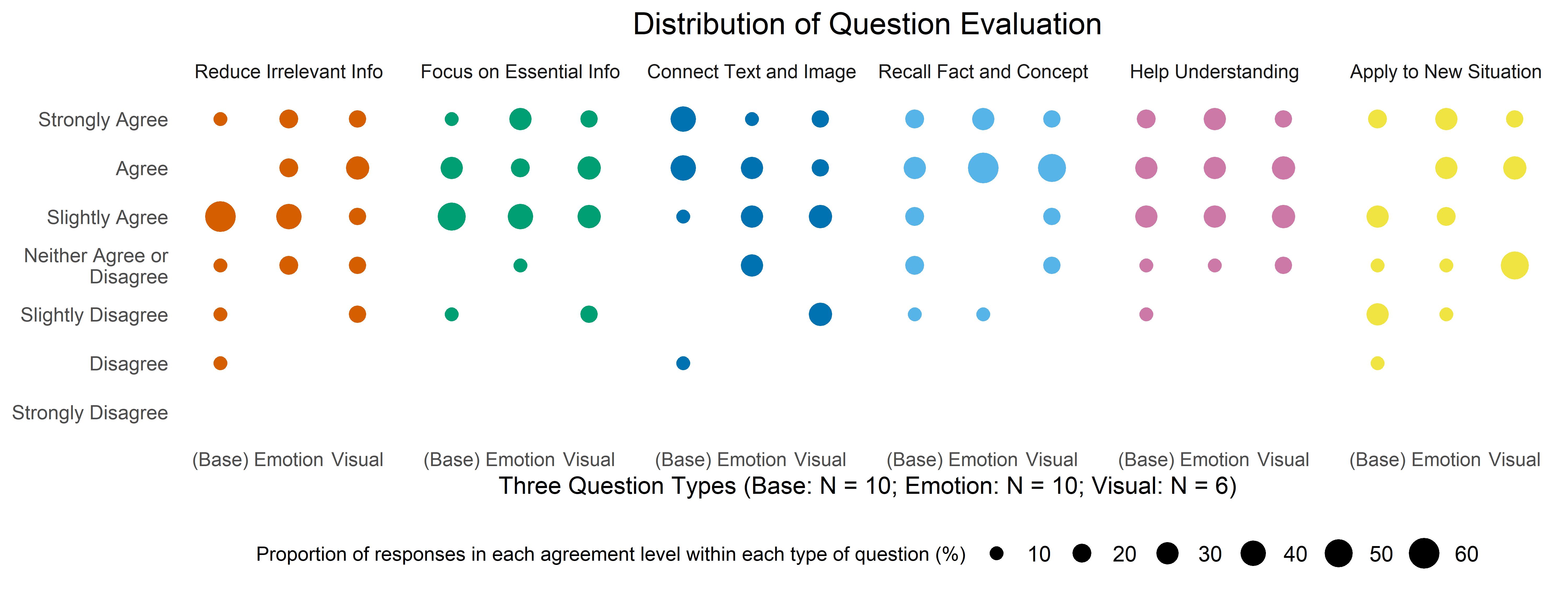}    
    \caption{DHH participants selected one, two, or all three question generation strategies in step 1 and evaluated their selected types after watching the video. Qualitatively, our two proposed question generation strategies (Emotion and Visual) seem to outscore the Base transcript strategy in recalling facts/concepts.  Emotion strategy scored higher for applying knowledge to new situations. Questions generated from the Base transcript received higher scores for fostering a connection between text and image.} 
    \Description{A series of scatter plots are aligned horizontally that shows the distribution of participants' agreement with different dimensions of question evaluation items. Participants had similar agreement levels with that the questions helped reduce irrelevant information in the learning across the three question generation strategies, with the larger dots located on the neutral to agreeing side showing that most participants agreed or held a neutral attitude to the statement. Each question generation strategy was agreed to be helpful by focusing on the essential information by most participants using it, with the large dots falling onto the agreeing side for this plot. The Base transcript questions attracted more participants agreeing that they helped connect the text and image than the Emotion and Visual questions, with the Emotion questions attracting a slightly larger portion of the participants agreeing that the questions were helpful from this perspective than the Visual questions did. The large dots for each question generation strategy that represent large portions of participants go from the agreeing side to the disagreeing side as the question generation strategies goes from the Base transcript questions, to the Emotion questions, then to the Visual questions in this plot. A larger portion of participants choosing the Emotion questions or the Visual questions agreed that the questions helped the learning by recalling facts and concepts than the participants choosing the Base transcript questions. The larger dots in the Emotion and Visual questions columns are on the agreeing side while small dots are evenly distributed from the agreeing side to the disagreeing side in the Base transcript questions column. There is no difference in the distribution of participants' agreement levels to the questions helping them better understand the knowledge across the three question generation strategies, with dots evenly distributed across different agreeing levels. Participants also had diverse views about the questions, helping them better apply the knowledge to new situations, and the views are similar across question types.}
    \label{fig:value1}
\end{figure}

In the comparison analysis of participants' choices of question generation strategies, we find 1) eight participants selected only one type with four selected Base Transcript, four for Emotion, and none for Visual) and 2) six participants selected Base + another question generation strategy OR Emotion + another question generation strategy OR Visual + another question generation strategy. 
The Visual Question generation strategy was selected only when another strategy was chosen simultaneously. Transcript strategy was perceived as Entrée, necessary for most learners; while Visual was associated with those who identify themselves with Deaf culture. 

\begin{figure}[ht]
    \centering  
    \includegraphics[width=1.0\textwidth]{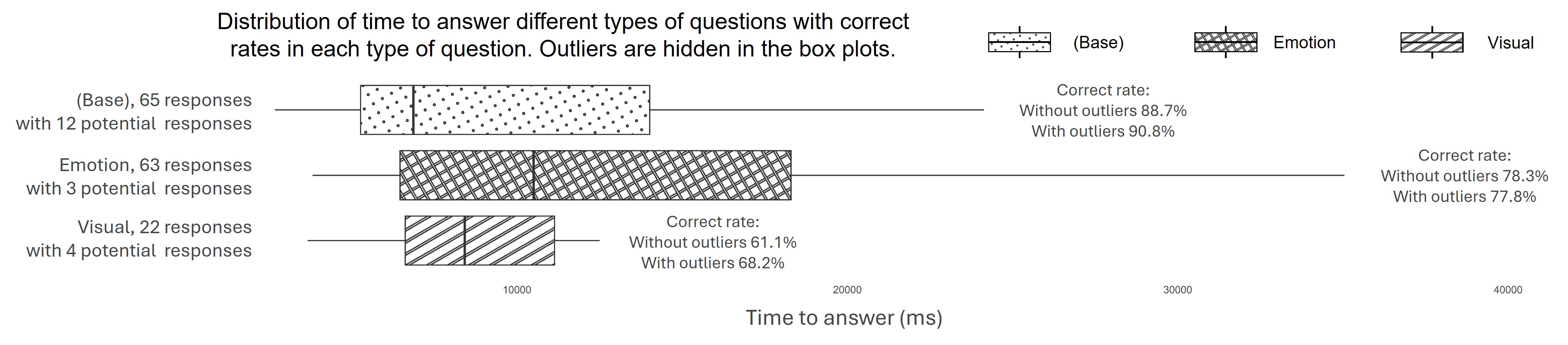}    
    \caption{Distribution of time to answer for different types of questions and the answering correct rates at the first attempts. After excluding potential outlier responses where participants took very long time to answer the questions, the results showed that the Base transcript questions took the shortest time to answer and received the highest correct rate, the Emotion Questions took the longest to answer and received an intermediate correct rate, and the Visual Questions took an intermediate time to answer with the lowest correct rate. The differences suggested that the Emotion and Visual Questions might identify video contents not fully understood by the participants, leading to longer answering time and lower correct rate. The Emotion Questions might elicit more thinking than the Visual Questions did, leading to a longer answering time and a higher correct rate.} 
    \Description{}
    \label{fig:tta}
\end{figure}

\textbf{Comparing Response Times and Accuracy.} For each question generation strategy, the time spent on answering a question and the correct rate at the first attempt were recorded and analyzed. The distribution of time-to-answer under each question generation strategy is demonstrated in the box plots in Fig. \ref{fig:tta}. The correct rate is denoted for each question generation strategy. Note: several extra long time-to-answer data points were treated as potential outliers and hidden from the figure displayed. Excluding the potential outliers, the Base transcript questions (based on 53 responses) had the shortest median (6,859 ms) and mean (7,215 ms) time-to-answer, the Visual Questions (based on 18 responses) had the second shortest median (8,414) and mean (7,938 ms) time-to-answer, while the Emotion Questions (based on 60 responses) took the longest time to complete (median: 10,500 ms; mean: 13,119 ms). For the correct rate, participants achieved the highest overall correct rate in the Base transcript questions (88.7\%), the lowest correct rate in the Visual Questions (61.1\%), and an intermediate correct rate in the Emotion Questions (78.3\%). The two question generation strategies with a lower answer correct rate and a longer time-to-answer may indicate that the Emotion and Visual Questions might associate with the video content that required participants to spend more time to comprehend and figure out the answer. Additionally, compared to the Emotion Questions, the Visual Questions took a shorter time to answer and received a lower correct rate. The difference might indicate that the Visual Questions did not elicit the amount of thinking that the Emotion Questions did, which might be the cause of the lower correct rate.

\textbf{Emotion Question Perceived to Improve Application of New Knowledge.} As shown in Fig. \ref{fig:value1}, participants perceived that Emotion Questions were helpful in almost all dimensions of learning, especially in Applying to New Situation--a higher order thinking skill based on Bloom's taxonomy. Participants considered Emotion Questions as beneficial they helped them see how others felt confused about the same content and raise their awareness to pay attention to the parts that might require more efforts to comprehend. As DHH learners, participants found it beneficial to learn with others from the same community who shared their language and culture, something that was rare in their previous online learning experience.

\begin{figure}[ht]
    \centering  
    \includegraphics[width=1.0\textwidth]{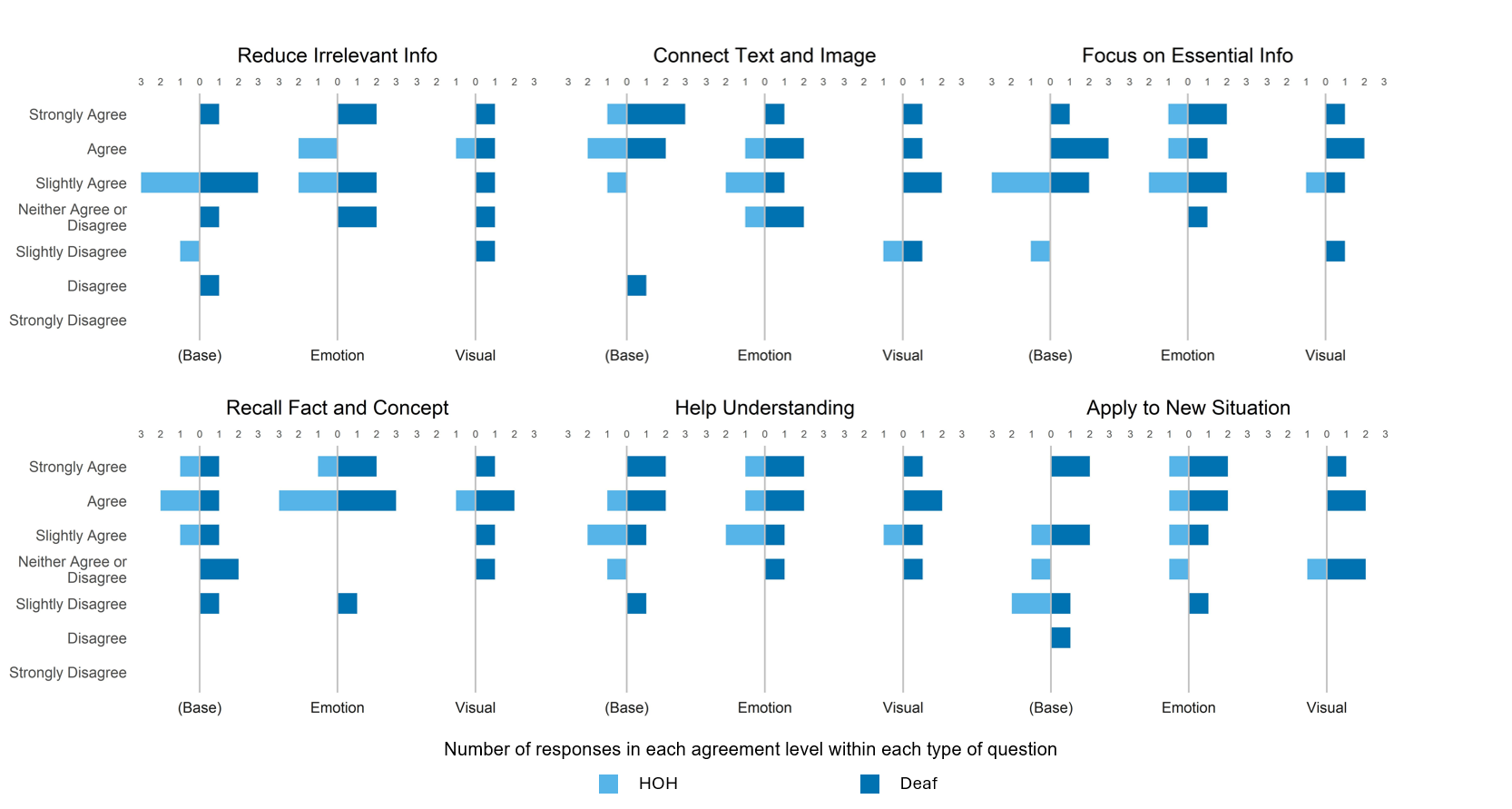}    
    \caption{Deaf (N=9) and HoH (N=6) participants rated the questions similarly in helping them learn, with some dimensions Deaf rating it more useful than HoH. Specially, Deaf participants rated pop-up questions in general more helpful for focusing on essential information and understanding the video than HoH. Deaf learners also rated Base transcript questions as most helpful for connecting text and image. Questions from the Visual Strategy were selected by five Deaf and only one HoH, indicating Deaf learners see more value in leveraging visual info in question generation to support learning using video beyond the basic transcript. The Visual Questions were rated similarly to Emotion in supporting learning. Also, the results indicate Deaf and HoH are two different learning populations to design with when it comes to video-based learning. } 
    \Description{}
    \label{fig:idrate}
\end{figure}

\textbf{Deaf Participants See More Value of Visual Questions.} As shown in Fig. \ref{fig:idrate}, five Deaf participants selected Visual but only one HoH learner selected the type. The mean of the Visual Questions rating was similar to the Emotion Questions rating. HoH participants did not select visuals and explained in the interview for the primary reason: they were unsure how generating questions on video timestamps with visual design problems could benefit their learning, especially given their fast reading speed, which allows them to switch quickly between captions and other information. As one participant (L1) noted: \textit{``I have no problem scanning between captions and the screen. I cannot speak for Deaf individuals, though I understand that lower caption reading speed might leave people with less time to process the visuals.''}

\textbf{Higher Ratings for Transcript Questions Associated with Higher Self-Efficacy.} By observing Fig. \ref{fig:value1}, Base transcript questions seem to be especially helpful for fostering a connection between text and image. The analysis further examined correlations between self-efficacy and participants' ratings for different question generation strategies. The data shows higher consensus on Base transcript questions as being helpful for focusing on essentials and connecting text and visuals which positively associated with some  self-efficacy scales.

Significant positive correlations were found between the question rating item for the Base transcript questions, that the questions helped learning by ``Focusing on Essential Information'', and self-efficacy evaluation items, that the participants were confident to perform different tasks more effectively ($r(8)$=.90, $p$<.01), and to perform well and improve performance in tough tasks ($r(8)$=.77, $p$<.01). There were significant positive associations between the Base transcript questions rating item, that the questions helped learning by ``Fostering Connection between Test and Image'', and self-efficacy evaluation items, that the participants could better accomplish difficult tasks ($r(8)$=.71, $p$=.02), do more tasks well compared to other people ($r(8)$=.71, $p$=.02), and perform well and improve performance in tough tasks ($r(8)$=.76, $p$=.01). In summary, participants who rated highly the Base transcript questions showed higher self-efficacy levels. At the same time, no significant findings for the other two question generation strategies. 

When exploring the correlations between the number of question generation strategies selected and the ratings for different strategies, the data shows that participants selecting the Base transcript + other question generation strategies rated the Base transcript questions higher than participants who only selected the Base transcript questions. The Point-biserial correlation test suggests that there were significant negative correlations between only selecting the Base transcript questions and the question ratings, that the questions helped learning by ``Reducing Irrelevant Information'' ($r(8)$=-.70, $p$=.02), and applying information in new situations not in this video ($r(8)$=-.70, $p$=.02). In summary, participants who selected the Base transcript + other question generation strategies showed higher valuation for the Base transcript questions than the participants who only chose the Base transcript questions.

In the interview, several participants noted that transcript questions helped them \textit{``make sure not to miss anything (in the captions)''}. However, these questions may focus on names and dates, prompting careful reading of captions and text, even though the information may not always be most important. Participants who rated transcript being very helpful noted in the interview that transcript was for \textit{``seeing how emotion and visual questions differ from basic questions''}, in other words, it was used to compare and critically assess different question generation strategies. 

\subsubsection{Suggestions for Personalized Questions in Support DHH Learners' Video-Based Learning}

\hspace{0in}

\textbf{Personalize Question Type based on Varied Language Background.} Feedback on the two types of questions (T/F and Multiple-choices in the study), varied among participants, reflecting diverse language preferences. Some participants expressed a preference for more T/F questions, citing the reduced reading load as a benefit. One participant noted, \textit{``I like the T/F questions. They’re a little less overwhelming than the ones with a lot of options.''} (D7) However, others found T/F questions tricky due to the nuances of English vocabulary, which created challenges in language comprehension. One participant even referred to T/F questions as ``traps.'' (D1) Specifically, missing a word or confusing word order and spelling was perceived as reasons for giving wrong answer on a T/F question, while such mistakes were less likely with multiple-choice questions.  Another participant also commented on a T/F question commenting \textit{``I possibly may be changing the wording around. I remember the first question. I was like really confused. And I don't know if it was, because it was the first question. I just remember I was like, I don't know, like the wording was just like a little bit weird and almost sounded like a bit of like a trick question.''} (D12) Another participant also commented on the T/F question as being language focused, \textit{``I thought it was an English question. Then I realized no it was not asking for my English.''} (D10)

Some participants felt that multiple-choice questions provided more context and background, enabling a more comprehensive learning experience. Others disliked multiple-choice questions as one participant remarked, \textit{“I know all the words. But all the answers have the same words. Hard to pick one.”} (D6) 

In addition to T/F and multiple-choice questions, some participants expressed a preference for more reciprocate and reflective questions that encourage critical thinking about the material, such as open-ended questions like “What do you think about this?.” 

To better accommodate diverse linguistic preferences, participants suggested offering options for simpler or more complex English language usage that correspond to personal choices when selecting  question types.  T/F questions are concise but easily misinterpreted, while multiple-choice can be overwhelming due to their length. 

\textbf{Personalizing Question Timing based on Varied Cognitive Focus Needs.} Participants indicated that the timing of question pop-ups could be personalized to enhance the learning experience. Some participants appreciated that the pop-ups appeared immediately after an important fact was introduced in the video and such a mechanism helped them stay focused and assess their understanding, while others felt that the question popups disrupted their flow of reading and diminished their sense of self-efficacy in overcoming learning challenges. One participant noted, \textit{``I noticed that the questions were about the material immediately explained in the video. That’s not challenging because it’s not hard to answer.''}(D13). Others found the question popups disruptive, with one explaining that ``\textit{the pop-ups often interrupt my flow,}'' and suggested introducing a clearer ``\textit{transition from learning mode to processing mode.}'' (D7)


Beyond the pop-up timing and content coverage of the questions, participants suggested additional feature enhancements for the prototype towards attention optimization. 
One participant indicated that adding visible markers on the video's progress bar could help learners anticipate when a question would appear and be prepared for it (D7). For long videos, options such as   ``you want less questions'', ``you want more questions'' could be provided to adapt to learners' needs for attention check-in (D10) .

To accommodate varied cognitive focus needs, the timing of question pop-ups can be personalized. For some participants, immediate pop-ups after a key fact is introduced help maintain engagement, while others find them less helpful as such popups caused distraction.  Additionally, some participants prefer the ability to control the pace of questions to better maintain their focus.

\textbf{Support Bilingual Interactions} Several participants recommended incorporating sign language in the learning videos and QA, as they felt that learning with their first language, i.e., sign language, than written English was more intuitive and beneficial. One participant expressed, \textit{``A lot of deaf people’s first language is not English. Can we have sign language support for deaf people who prefer sign language?''} (D10). Instead of suggesting to have the prototypes in ASL-only, most participants preferred to have a video-based learning system with bilingual capability that supports the use of both ASL, a full-fledged natural language with its own grammar and lexicon, and English.  
Other than video lectures and QA with sign language support, participants suggested adding visual clues to enhance the readability of the video captions and the questions. One participant wondered \textit{``Maybe can make some words bold? So easier to read?''} (D7) The interviews emphasized the need for all English content and interactions to be available in ASL, the preferred language for a large proportion of the DHH community. 

\textbf{Summary of RQ1} Our proposed prototype was found beneficial by DHH participants, with the two proposed question generation strategies offering additional benefits beyond the Base transcript questions, such as Deaf participants placed more value on Visual Questions. Participants suggested personalization features to further improve learning accessibility of the DHH community, particularly by considering language diversity within the community and individuals' cognition focus needs when watching learning videos.

\subsection{RQ2-Challenges Faced by DHH Learners in LLM-Powered Interactions}
This section uses interview data collected in Step 4, and attitude towards ChatGPT collected in Step 1 (pre-learning) and Step 4 (post-learning) to answer RQ2. 

\subsubsection{English-only Conversation Wasn't Always Understandable} 
As previously noted, English-only interaction with the chatbot embedded in the prototype proved insufficient for DHH participants, and they suggested the need for additional support in sign language or visual modalities. This section delves further into the specific challenges participants faced when engaging with a LLM-powered chatbot, emphasizing the importance of addressing language diversity in chatbot interaction design.

In Step 1 of the study, participants were instructed to independently engage in a text-based conversation with ChatGPT to select question generation strategies. While researchers were supposed to intervene only when participants encountered significant difficulty, most participants sought assistance frequently. Four participants bypassed the chatbot’s self-introduction (Fig. \ref{fig:system_screens} in green text) and immediately asked researchers for help. Three participants ignored the chatbot's prompt to select a strategy and instead, typed their names. Two participants engaged in unrelated conversations (e.g., asking about Olympic games) before seeking guidance. This pattern suggests that participants found the chatbot's text-based instructions unclear or difficult to follow.  Below presents a sample conversation of L2 with the chatbot in Step 1:
{\fontfamily{qcr}\selectfont
\begin{quote}
\begin{itemize} 

\item Chatbot: "Let's go over the three types of questions:

1. **Transcript-based:**
[...]

2. **Transcript+Emotion Cue-based:**
  [...]

3. **Transcript+Visual Cue-based:**
   [...]

Would you like to choose one (1), two (2), or all three (3) types of questions? Please let me know your choice."

\item Participant: ``yes three questions ''

\item Chatbot: "Great! You have chosen all three types of questions. Let's confirm:

1. Transcript-based
2. Transcript+Emotion Cue-based
3. Transcript+Visual Cue-based

Is this correct? If yes, I will proceed with generating the questions." 

\item Participant: ``3''

\item Chatbot: "Got it! You have chosen two types of questions:

1. Transcript-based
2. Transcript+Emotion Cue-based

Is this correct? If yes, I will proceed with generating the questions.'' 

\item Participant: ``2''

...
\end{itemize}
\end{quote} 
}

Approximately half of the participants requested clarification from the researchers regarding what to type, reflecting the inadequacy of the chatbot's instructions. Although most participants eventually moved on to Step 2, they often expressed uncertainty about their understanding of the English prompts. One participant remarked, ``\textit{Sometimes I read the English sentences, I thought I understood but maybe I didn't}' (D6). Additionally, seven participants did not respond to any chatbot prompted questions so the intended interaction/conversation with the chatbot abruptly ended. Three participants did not correctly select their intended question generation strategy. They aimed to choose the third option by typing “3” or “three,” but inadvertently selected all three strategies, and vice versa. Although the prototype was programmed to explicitly confirm their selection by saying, “Do you want strategy xx, xx, and xx?”, the participants did not correct the chatbot and proceeded with an unintended choice. Additionally, none of the participants made language-specific requests to the chatbot, such as ``please be precise,'' as shown in the UI sample screenshot Fig. \ref{fig:system_screens}. 

These observations demonstrated that English-only interaction was not only insufficient but also led to participant confusion or moments of disengagement with the chatbot prompts. The participants suggested some interactions with LLM should be more ``visual'' to reduce the possible miscommunication when only text-based information/prompts are used. For example, the UI can include checklists where learners can interact without ambiguity, ``\textit{About the chatbot in the beginning, I was expecting to see a drop-down menu or something like that. So I could choose the option. Was not sure what I should type. }''(D10). Admittedly, many DHH learners might have not used ChatGPT or equivalent tools before and felt uneasy interacting with the chatbot. But the absence of sign language or missing visual support may have compounded the challenges. The goal of fostering independent learning was undermined by frequent reliance on researcher intervention.

Some participants attributed the miscommunication to the chatbot's lengthy and unfocused responses. In the conversation snippet from L12, the chatbot used formats such as bolding (**), bullet points (-), and numbering to differentiate content, but participants still found it difficult to focus on key information when skimming through the text. As D15 reflected on his interaction, he noted, ``\textit{I am not sure how that info slipped my mind, I just skim through and nothing caught my eye. And I can tell this is ChatGPT because of all the formatting, maybe that's what I always do when dealing with LLM. I intern at *** [a large tech company] so I am quite familiar with all the LLM things. I think maybe bolding the question or moving the question to the topic might be more readable. Formatting issues.}''

These findings revealed several challenges for DHH participants interacting directly with a LLM-powered interface. First, uncertainty about how to interact with the tool by asking for clarification or requesting alternative expressions, even when they didn’t fully understand the chatbot’s prompts (lack of training in addition to the language issue). Second, semantic issues were noted, such as confusion around words like "second" or “third,” which were not clearly communicated. Third, the format and length of the chatbot’s responses made it cognitively challenging to follow, with some participants finding it difficult to engage with.

The DHH participants might not have fully understood the level of control they had over the chatbot’s responses or lacked the confidence to adjust the interaction for better clarity. The absence of participant interaction with the LLM enabled tool indicated two important issues: 1) a need for educators to provide the necessary training for their DHH students about ChatGPT or its equivalent tools and how to use them so these special students are not at a disadvantage with their hearing peers and 2) a need for the system to better guide learners on how to tailor the chatbot’s communication style to their preferences. This is particularly important considering the linguistic diversity of the DHH community. As reviewed in related works, DHH individuals are considered bilingual because a large percentage of them use both a sign language and a spoken language. For bilingual speakers, code-switching is a phenomenon that happens when a speaker alternates between two or more languages during a single conversation, sentence, or situation. Therefore, DHH participants have to constantly switch between two languages, which may result in misunderstandings during English-only interactions with the LLM-powered chatbot. For instance, one participant, D14 typed to the chatbot: \textit{I am interesting in web developement, which would you recommend language should I learn?} This further emphasizes the need for chatbots to be more linguistically flexible when interacting with DHH learners.

\begin{figure}[ht]
    \centering  
    \includegraphics[width=1.05\textwidth]{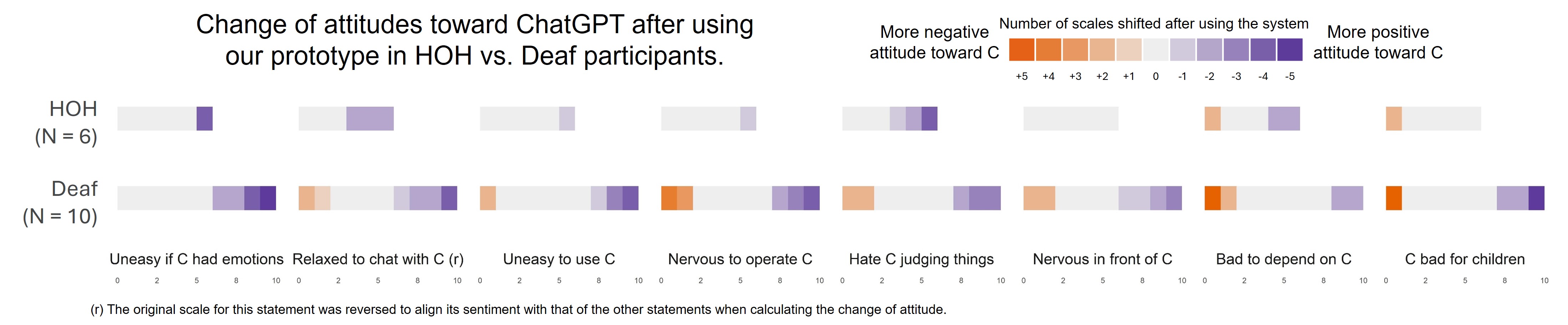}    
    \caption{HoH participants (N=6) reported minimal changes in their attitudes towards the ChatGPT-powered experience before and after using our prototype. Deaf participants (N=10) exhibited both more positive and negative shifts in attitude. This indicates that our prototype has impacted participants' attitudes toward ChatGPT. Note: In the legend, colors further to the right indicate stronger disagreement with the statement below, while the opposite is true for the one statement marked 'r.' The order of colors in the legend corresponds to the sequence they appear in the agreement bar for each statement above. Each block represents one participant's response. 
    } 
    \Description{}
    \label{fig:attitude}
\end{figure}

\subsubsection{Peer Influence and Internalized Caution towards Rapidly Evolving LLM} 
Survey results show that DHH participants' attitudes towards ChatGPT changed by using our prototype. The attitudes were measured by 7-point Likert Scales about the participants' agreement level with a series of statements with negative sentiment towards ChatGPT. Therefore, a lower value indicates a more positive attitude towards ChatGPT, and a negative change in the value indicates a change towards a more positive attitude towards ChatGPT. The changes in attitude among Deaf and HoH individuals are visualized in Fig. \ref{fig:attitude}. Due to the relatively small size of the experimental sample, no statistical test was performed, and only the descriptive statistics is presented. HoH participants (N=6) reported minimal changes in their attitudes towards the ChatGPT before and after using our prototype  (pre-learning M = 3.38, SD = 0.81; post-learning M = 2.98, SD = 0.70; changes  M = -0.39, SD = 0.35). In contrast, Deaf participants (N=10) exhibited both more positive and negative shifts in attitude (pre-learning M = 3.95, SD = 0.44; post-learning M = 3.51, SD = 0.78; changes  M = -0.44, SD = 0.72).  

Pre-study interviews revealed that all participants had used  ChatGPT. Seven reported using it daily, while four used it several times a week. Four had rarely used it before and do not use it now. 
Participants primarily engaged with ChatGPT for grammar checks, idea generation, and answering questions about unfamiliar topics. D5 found the rapid evolution of the technology overwhelming, as they worked hard to keep up with the changes and figure out how to effectively work with it, and our prototype \textit{``I will generally understand (ChatGPT), because if you give it too much information, AI gets confused, but I wouldn't say I had a hard time. I would say I just had to sort of learn how to do it all the time. The quality, I think, depends a lot on what you feed into it. How you say, what you type into chatGPT.''}

In the post-study interviews, participants’ perceptions of ChatGPT shifted, aligning with the survey results. Many were encouraged and inspired by learning how ChatGPT could be used in specific, positive ways to benefit the DHH community. For instance, D10 expressed ``\textit{People feel like AI can steal their jobs. Deaf communities in general lacking of resources. I feel I learned more about what AI can do (today). Never really thought AI could be applied in situations like this... This experiment helps me understand better what AI can help us with. One example would be when you read a book, AI can help you read better. Sometimes I read English sentences, and I thought I understood but maybe I didn't. So this technology can be used in situations like this. I can see this technology being helpful with language learning as well. Does not have to be limited to video only. }''

The post-learning interviews were not able to fully clarify why there were eight participants' responses that became more negative on at least one scale. An analysis of the interviews suggests that participants' perceptions were influenced by how their DHH peers used the tool and perceived it, and the study provided an opportunity to probe individuals' thoughts about ChatGPT and the perceptions within the community, whether positive or negative. The phenomenon about deaf people's need for ingroup identification or conformity \cite{carter2015deaf} was manifested in the participants’ strong connection to their DHH peers' views on ChatGPT, leading some to align their opinions with those of their peers in the group. For instance, D16, a student from an institution for deaf shared her thoughts in pre-learning, \textit{“I don’t use it as much; my friends use it a lot, and she said the tools generate good language, so I have used it for generating cover letters once, I think it is okay”} and in the post-learning interview, she expressed, \textit{“ChatGPT does generate good language, and I want to make sure I don’t rely too much on it. I am not sure if the Deaf community can all use ChatGPT-related technology in good ways or think it is beneficial. ”} This indicates that social conformity has played its role, as participants reflected on the experiences of others in the DHH community when forming their own post-study opinions.


\subsubsection{Varied Trust in LLM to be Objective or Inclusive} Some participants chose the Base transcripts only, concerned that DHH identity might be inaccurately represented in the LLM based on stereotypes. Unlike the Emotion and Visual types, the Base transcript option maintains the language authenticity, so selecting the transcript generation strategy may mitigate issues with the LLM incorrectly interpreting DHH identity and its associated learning/language preferences. Here is a quote from a participant who expressed less trust in whether LLM truly understands DHH learners' preferences: ``\textit{Where did you get the other people's data? ...[researcher explaining process]... Oh really? I thought you told GPT to generate content for deaf students directly. Real data is better than directly telling GPT. GPT is biased. }''

When participants expressed interest in finding where the DHH related visual or emotion data came from , and how the data was fed into ChatGPT, the researcher would explain the process and \textit{how} DHH related data were learner-sourced in question generation. Participants noted that the learner-sourcing process was too complex to be fully understood, and it was challenging to trust the technology can \textit{access and incorporate} DHH data in correct ways.  In summary, explaining how LLM incorporates DHH data to DHH learners to gain their trust in the technology is important yet challenging.

\subsubsection{Varied Views on Whether ``DHH Option'' is Good For DHH Learning Effectiveness} 

Some participants asked the researchers if the Emotion and Visual questions were easier versions of the Base transcript. By 'easier,' some were interested in whether it meant 'language only' easier with reduced language complexity, or if it was 'content-wise' easier with a focus on lower levels of thinking skills, such as recall versus reflection questions. Three participants were concerned that the 'DHH options' might be like feeding candy to a child, tricking learners into selecting easier options and avoiding challenges. While this may be acceptable in other settings, it is not as beneficial in a learning environment.

One participant, majoring in government and political science, brought up a very interesting point for educators to ponder. She stated that as a HoH individual who grew up in both mainstream and special education schools, she felt that mainstream education made her feel more prepared for the post-graduation world, where her work performance would be evaluated similarly to that of hearing individuals. In comparison, special education schools had significantly lower writing and reading workloads, which made her feel less prepared for post-graduation and more nervous about working with hearing individuals. She pinpointed the broader goal of special education for people with disabilities and its current practices: \textit{``Well, I feel we have too few writing and reading assignments at special education schools, and no AI-related content is introduced in classrooms. I believe this puts us at a higher risk of not being prepared to succeed after we leave school. This is a very complicated question. Since the IDEA law in 1990, which made schools responsible for providing interpreters, we've seen more and more Deaf kids entering the mainstream. Does this suggest that mainstream education, not just Deaf kids, better prepares kids for work? Is current special education being left behind? I keep thinking about these questions and motivate me to get a PhD.''} (D16)

\section{Discussions}








\subsection{Automatic Generation Pipeline}

In this study, the research team studied the potential of using automatically generated questions to enhance personalized learning experiences for DHH learners. Although the questions were pre-generated before DHH participants watching the video, the question co-generation pipeline (as shown in Fig. \ref{fig:question_gen}) is extensible to enable on-the-fly question generation during the learning process following the prompt engineering process described in the study. 

Learners can choose when and where to receive a pop-up question and customize prompts to generate diverse questions that meet their expectations. Questions of varying types, difficulty levels, and language styles can be generated to enhance learners experience. However, generating questions using LLMs via prompts and evaluating them can be challenging, as it requires users to have the knowledge and skills to work effectively with LLMs~\cite{zamfirescu2023johnny}. Therefore, before DHH learners can create personalized question content using LLMs for video-based learning, further research is needed to understand how to foster DHH learners' ability to interact with LLMs for tailored content generation. 


As the question generation pipeline becomes more extensible, it can leverage learner feedback to improve the question bank. Beyond real-time question generation, the pipeline can create a data loop where DHH learners’ reactions and ratings serve as feedback for future model iterations~\cite{wu2022survey}. The LLM can refine questions and adjust question difficulty based on learner performance and profiles. Furthermore, if learners can customize questions, peer-sourced questions will accumulate, enriching the resource pool for the DHH community. Such DHH learner data can also be used for post-training the model, allowing the LLM to better adapt to DHH learners' needs.

In summary, this study serves as a starting point for leveraging LLMs to enhance DHH learners' learning experiences. The co-generation pipeline has the potential to extend into a fully automated generation loop, aligning the LLM with the specific learning needs of DHH learners. 

\subsection{Special Education Instructor's Blind Spot: Lived Experiences of Learners with Disabilities} RQ1's findings offer valuable insights into quiz question design for accessible learning. While instructors prioritized questions that could be understood by the largest possible DHH learner population, some learners expressed a desire for more personalized approaches. Although the mismatch between instructor and learner preferences has been studied before \cite{wang2021seeing}, this study shows that similar issue also persists among experienced instructors in special education. Despite the use of personalized approaches, such as smaller class sizes, for DHH learners, specifically, personalization is strongly connected to language diversity and unique challenges in processing multimedia content. Although this study did not include DeafBlind users, some participants were curious about how questions might be generated for such users. Since accessibility is a lived experience and it is difficult for instructors to fully anticipate each individual's needs, further personalizing special education experiences through technology that empowers both instructors and learners becomes increasingly essential.

\subsection{Possibilities in Learner-sourcing DHH Learners' Data}This study utilized two types of data from DHH learners to generate quiz questions: collective emotional responses for emotion-based questions and visual questions derived from synthesized learning preferences of Deaf learners. Due to the limited number of participants, a quantitative comparison of different generation strategies was not feasible. However, qualitative results suggest that the two strategies enhanced certain aspects of the learning experience (e.g., participant ratings) and led to variations in question response times and accuracy rates compared to the more traditional Base transcript method. While this study explored emotion and visual data as two types of learner data that inform LLMs about DHH learning preferences, these are only examples based on small datasets. Further data collection is essential to make future research more robust in studying additional types of learning data that can enhance personalization and accessible learning. 

Challenges identified in RQ2 directly affect the feasibility of addressing RQ1. DHH learners exhibited varying levels of trust in the inclusivity of LLMs and expressed diverse views on whether the 'DHH option' improves learning outcomes. For instance, concerns persist about whether LLMs can accurately incorporate DHH data without perpetuating existing stereotypes. Moreover, significant challenges remain regarding how DHH users' data is used and communicated to the community. Lastly, the study did not examine whether certain types of learning data may present greater sensitivity or privacy risks during collection and analysis. Future research must tackle these issues to ensure that personalized learning technologies are both ethical and effective in meeting the diverse needs of learners.  Additionally, researchers should be prepared to face similar issues with limited number of participants, which may affect the feasibility of quantitative analysis.

\subsection{Methodological Reflections} RQ2 revealed that interacting with the prototype led to both positive and negative changes in participants' attitudes towards ChatGPT. Most within-subject changes collected via survey exceeded one scale point, indicating these shifts were less likely to be random. It was particularly interesting to observe how just one experimenting prototype could influence attitudes, enabling participants to recognize ChatGPT’s potential for serving the DHH community. This swift change in perspective suggests a need for more tools to enhance ChatGPT experiences and training of proper use of these tools so independent and objective thinking about ChatGPT can be fostered among DHH learners. Conversely, researchers should be mindful of how new LLM-based prototypes and studies are introduced, as these interactions can impact DHH learners' perceptions about ChatGPT and potentially the LLM technology. The potential influence of these experiences underscores the importance of ethical and reflective practices in developing and testing AI tools within minority communities.

\section{Conclusion}

We conducted a study with 16 DHH participants to explore the use of LLMs in generating quiz questions based on previous DHH learners' emotional responses and visual learning preferences. While the results showed that the proposed two strategies improved some dimensions of the learning experience compared to traditional transcript-based methods, several challenges and needs were identified. DHH learners with language diversity have additional barriers when interacting with text-based LLM-empowered tools. Lack of training and less proficiency in written English were factors that could hinder their ability to benefit from recent LLM advancements fully.
Additionally, learners expressed varying levels of trust in the inclusivity of LLM-generated content, with concerns about potential system/language bias as the models were trained with written English and the transparency of how DHH-related data is used. Furthermore, an unmet need for proper training to use the LLM tools was echoed by study participants. Thus, further research is needed to 1) refine LLMs to include robust datasets collected from special population with diverse linguistic needs, such as DHH individuals and other people with English as a second or foreign language and 2) improve interaction design or prompt engineering with LLMs so it can better accommodate users with diverse backgrounds and achieve improved accessibility for all. 
Overall, this study underscores the potential of LLM to enhance personalized learning for DHH learners but also highlights the importance of addressing new ethical and technical barriers inherent in the advancement and adoption of AI.

\bibliographystyle{ACM-Reference-Format}

\begin{thebibliography}{65}


\ifx \showCODEN    \undefined \def \showCODEN     #1{\unskip}     \fi
\ifx \showDOI      \undefined \def \showDOI       #1{#1}\fi
\ifx \showISBNx    \undefined \def \showISBNx     #1{\unskip}     \fi
\ifx \showISBNxiii \undefined \def \showISBNxiii  #1{\unskip}     \fi
\ifx \showISSN     \undefined \def \showISSN      #1{\unskip}     \fi
\ifx \showLCCN     \undefined \def \showLCCN      #1{\unskip}     \fi
\ifx \shownote     \undefined \def \shownote      #1{#1}          \fi
\ifx \showarticletitle \undefined \def \showarticletitle #1{#1}   \fi
\ifx \showURL      \undefined \def \showURL       {\relax}        \fi
\providecommand\bibfield[2]{#2}
\providecommand\bibinfo[2]{#2}
\providecommand\natexlab[1]{#1}
\providecommand\showeprint[2][]{arXiv:#2}

\bibitem[Agrawal and Peiris(2021)]%
        {agrawal2021eyetracking}
\bibfield{author}{\bibinfo{person}{Chanchal Agrawal} {and} \bibinfo{person}{Roshan~L Peiris}.} \bibinfo{year}{2021}\natexlab{}.
\newblock \showarticletitle{I See What You’re Saying: A Literature Review of Eye Tracking Research in Communication of Deaf or Hard of Hearing Users}. In \bibinfo{booktitle}{\emph{Proceedings of the 23rd International ACM SIGACCESS Conference on Computers and Accessibility}} (Virtual Event, USA) \emph{(\bibinfo{series}{ASSETS '21})}. \bibinfo{publisher}{Association for Computing Machinery}, \bibinfo{address}{New York, NY, USA}, Article \bibinfo{articleno}{41}, \bibinfo{numpages}{13}~pages.
\newblock
\showISBNx{9781450383066}
\urldef\tempurl%
\url{https://doi.org/10.1145/3441852.3471209}
\showDOI{\tempurl}


\bibitem[Alnahdi et~al\mbox{.}(2024)]%
        {alnahdi2024enhancing}
\bibfield{author}{\bibinfo{person}{Ghaleb~H Alnahdi}, \bibinfo{person}{Arwa Alwadei}, {and} \bibinfo{person}{Nuha Alharbi}.} \bibinfo{year}{2024}\natexlab{}.
\newblock \showarticletitle{Enhancing special education programs’ curricula for students with intellectual disabilities in saudi arabia: A call for personalized approaches and inclusive practices}.
\newblock \bibinfo{journal}{\emph{Research in Developmental Disabilities}}  \bibinfo{volume}{151} (\bibinfo{year}{2024}), \bibinfo{pages}{104785}.
\newblock


\bibitem[Alrashidi(2023)]%
        {alrashidi2023synergistic}
\bibfield{author}{\bibinfo{person}{Malek Alrashidi}.} \bibinfo{year}{2023}\natexlab{}.
\newblock \showarticletitle{Synergistic integration between internet of things and augmented reality technologies for deaf persons in e-learning platform}.
\newblock \bibinfo{journal}{\emph{The Journal of Supercomputing}} \bibinfo{volume}{79}, \bibinfo{number}{10} (\bibinfo{date}{July} \bibinfo{year}{2023}), \bibinfo{pages}{10747--10773}.
\newblock
\showISSN{1573-0484}
\urldef\tempurl%
\url{https://doi.org/10.1007/s11227-022-04952-z}
\showDOI{\tempurl}


\bibitem[AlShawabkeh et~al\mbox{.}(2023)]%
        {alshawabkeh2024covid}
\bibfield{author}{\bibinfo{person}{Abdallah AlShawabkeh}, \bibinfo{person}{Faten Kharbat}, \bibinfo{person}{Ajayeb~Abu Daabes}, {and} \bibinfo{person}{M~Lynn Woolsey}.} \bibinfo{year}{2023}\natexlab{}.
\newblock \showarticletitle{Technology-based Learning and the Digital Divide for Deaf/Hearing Students During Covid-19: Academic Justice Lens in Higher Education}.
\newblock \bibinfo{journal}{\emph{Educational Technology \& Society}} \bibinfo{volume}{26}, \bibinfo{number}{4} (\bibinfo{year}{2023}), \bibinfo{pages}{pp. 136--149}.
\newblock
\showISSN{11763647, 14364522}
\urldef\tempurl%
\url{https://www.jstor.org/stable/48747526}
\showURL{%
\tempurl}


\bibitem[Arif et~al\mbox{.}(2024)]%
        {arif2024generation}
\bibfield{author}{\bibinfo{person}{Taimoor Arif}, \bibinfo{person}{Sumit Asthana}, {and} \bibinfo{person}{Kevyn Collins-Thompson}.} \bibinfo{year}{2024}\natexlab{}.
\newblock \showarticletitle{Generation and Assessment of Multiple-Choice Questions from Video Transcripts using Large Language Models}. In \bibinfo{booktitle}{\emph{Proceedings of the Eleventh ACM Conference on Learning@ Scale}}. \bibinfo{pages}{530--534}.
\newblock


\bibitem[Banda et~al\mbox{.}(2011)]%
        {banda2011review}
\bibfield{author}{\bibinfo{person}{Devender~R Banda}, \bibinfo{person}{Maud~S Dogoe}, {and} \bibinfo{person}{Rose~Marie Matuszny}.} \bibinfo{year}{2011}\natexlab{}.
\newblock \showarticletitle{Review of video prompting studies with persons with developmental disabilities}.
\newblock \bibinfo{journal}{\emph{Education and Training in Autism and Developmental Disabilities}} (\bibinfo{year}{2011}), \bibinfo{pages}{514--527}.
\newblock


\bibitem[Bandura and Wessels(1997)]%
        {bandura1997self}
\bibfield{author}{\bibinfo{person}{Albert Bandura} {and} \bibinfo{person}{Sebastian Wessels}.} \bibinfo{year}{1997}\natexlab{}.
\newblock \bibinfo{booktitle}{\emph{Self-efficacy}}.
\newblock \bibinfo{publisher}{Cambridge University Press Cambridge}.
\newblock


\bibitem[Bat-Chava(2000)]%
        {Bat-Chava2000diversity}
\bibfield{author}{\bibinfo{person}{Yael Bat-Chava}.} \bibinfo{year}{2000}\natexlab{}.
\newblock \showarticletitle{Diversity of Deaf Identities}.
\newblock \bibinfo{journal}{\emph{American Annals of the Deaf}} \bibinfo{volume}{145}, \bibinfo{number}{5} (\bibinfo{year}{2000}), \bibinfo{pages}{420--428}.
\newblock
\urldef\tempurl%
\url{http://www.jstor.org/stable/44393234}
\showURL{%
\tempurl}
\newblock
\shownote{Accessed 31 Aug. 2024}.


\bibitem[Bhavya et~al\mbox{.}(2022)]%
        {bhavya2022collaborativecaption}
\bibfield{author}{\bibinfo{person}{Bhavya Bhavya}, \bibinfo{person}{Si Chen}, \bibinfo{person}{Zhilin Zhang}, \bibinfo{person}{Tiffany Li}, \bibinfo{person}{Chengxiang Zhai}, \bibinfo{person}{Lawrence Angrave}, {and} \bibinfo{person}{Yun Huang}.} \bibinfo{year}{2022}\natexlab{}.
\newblock \showarticletitle{Exploring collaborative caption editing to augment video-based learning}.
\newblock \bibinfo{journal}{\emph{Educational technology research and development}}  \bibinfo{volume}{70} (\bibinfo{date}{07} \bibinfo{year}{2022}).
\newblock
\urldef\tempurl%
\url{https://doi.org/10.1007/s11423-022-10137-5}
\showDOI{\tempurl}


\bibitem[Bragg et~al\mbox{.}(2021)]%
        {bragg2021fate}
\bibfield{author}{\bibinfo{person}{Danielle Bragg}, \bibinfo{person}{Naomi Caselli}, \bibinfo{person}{Julie~A. Hochgesang}, \bibinfo{person}{Matt Huenerfauth}, \bibinfo{person}{Leah Katz-Hernandez}, \bibinfo{person}{Oscar Koller}, \bibinfo{person}{Raja Kushalnagar}, \bibinfo{person}{Christian Vogler}, {and} \bibinfo{person}{Richard~E. Ladner}.} \bibinfo{year}{2021}\natexlab{}.
\newblock \showarticletitle{The FATE Landscape of Sign Language AI Datasets: An Interdisciplinary Perspective}.
\newblock \bibinfo{journal}{\emph{ACM Trans. Access. Comput.}} \bibinfo{volume}{14}, \bibinfo{number}{2}, Article \bibinfo{articleno}{7} (\bibinfo{date}{jul} \bibinfo{year}{2021}), \bibinfo{numpages}{45}~pages.
\newblock
\showISSN{1936-7228}
\urldef\tempurl%
\url{https://doi.org/10.1145/3436996}
\showDOI{\tempurl}


\bibitem[Braun and Clarke(2012)]%
        {braun2012thematic}
\bibfield{author}{\bibinfo{person}{Virginia Braun} {and} \bibinfo{person}{Victoria Clarke}.} \bibinfo{year}{2012}\natexlab{}.
\newblock \bibinfo{booktitle}{\emph{Thematic analysis.}}
\newblock \bibinfo{publisher}{American Psychological Association}.
\newblock


\bibitem[Butler et~al\mbox{.}(2019)]%
        {butler2019speechrecognition}
\bibfield{author}{\bibinfo{person}{Janine Butler}, \bibinfo{person}{Brian Trager}, {and} \bibinfo{person}{Byron Behm}.} \bibinfo{year}{2019}\natexlab{}.
\newblock \showarticletitle{Exploration of Automatic Speech Recognition for Deaf and Hard of Hearing Students in Higher Education Classes}. In \bibinfo{booktitle}{\emph{Proceedings of the 21st International ACM SIGACCESS Conference on Computers and Accessibility}} (Pittsburgh, PA, USA) \emph{(\bibinfo{series}{ASSETS '19})}. \bibinfo{publisher}{Association for Computing Machinery}, \bibinfo{address}{New York, NY, USA}, \bibinfo{pages}{32–42}.
\newblock
\showISBNx{9781450366762}
\urldef\tempurl%
\url{https://doi.org/10.1145/3308561.3353772}
\showDOI{\tempurl}


\bibitem[Callender and McDaniel(2007)]%
        {callender2007benefits}
\bibfield{author}{\bibinfo{person}{Aimee~A Callender} {and} \bibinfo{person}{Mark~A McDaniel}.} \bibinfo{year}{2007}\natexlab{}.
\newblock \showarticletitle{The benefits of embedded question adjuncts for low and high structure builders.}
\newblock \bibinfo{journal}{\emph{Journal of Educational Psychology}} \bibinfo{volume}{99}, \bibinfo{number}{2} (\bibinfo{year}{2007}), \bibinfo{pages}{339}.
\newblock


\bibitem[Carter(2015)]%
        {carter2015deaf}
\bibfield{author}{\bibinfo{person}{Michael~J Carter}.} \bibinfo{year}{2015}\natexlab{}.
\newblock \showarticletitle{Deaf identity centrality: Measurement, influences, and outcomes}.
\newblock \bibinfo{journal}{\emph{Identity}} \bibinfo{volume}{15}, \bibinfo{number}{2} (\bibinfo{year}{2015}), \bibinfo{pages}{146--172}.
\newblock


\bibitem[Chen et~al\mbox{.}(2024b)]%
        {chen2024gptutor}
\bibfield{author}{\bibinfo{person}{Eason Chen}, \bibinfo{person}{Jia-En Lee}, \bibinfo{person}{Jionghao Lin}, {and} \bibinfo{person}{Kenneth Koedinger}.} \bibinfo{year}{2024}\natexlab{b}.
\newblock \showarticletitle{GPTutor: Great Personalized Tutor with Large Language Models for Personalized Learning Content Generation}. In \bibinfo{booktitle}{\emph{Proceedings of the Eleventh ACM Conference on Learning@ Scale}}. \bibinfo{pages}{539--541}.
\newblock


\bibitem[Chen et~al\mbox{.}(2024a)]%
        {chen2024signmaku}
\bibfield{author}{\bibinfo{person}{Si Chen}, \bibinfo{person}{Haocong Cheng}, \bibinfo{person}{Jason Situ}, \bibinfo{person}{Desir\'{e}e Kirst}, \bibinfo{person}{Suzy Su}, \bibinfo{person}{Saumya Malhotra}, \bibinfo{person}{Lawrence Angrave}, \bibinfo{person}{Qi Wang}, {and} \bibinfo{person}{Yun Huang}.} \bibinfo{year}{2024}\natexlab{a}.
\newblock \showarticletitle{Towards Inclusive Video Commenting: Introducing Signmaku for the Deaf and Hard-of-Hearing}. In \bibinfo{booktitle}{\emph{Proceedings of the CHI Conference on Human Factors in Computing Systems}} (Honolulu, HI, USA) \emph{(\bibinfo{series}{CHI '24})}. \bibinfo{publisher}{Association for Computing Machinery}, \bibinfo{address}{New York, NY, USA}, Article \bibinfo{articleno}{56}, \bibinfo{numpages}{18}~pages.
\newblock
\showISBNx{9798400703300}
\urldef\tempurl%
\url{https://doi.org/10.1145/3613904.3642287}
\showDOI{\tempurl}


\bibitem[Chen et~al\mbox{.}(2023)]%
        {chen2023thinkaloud}
\bibfield{author}{\bibinfo{person}{Si Chen}, \bibinfo{person}{Desir\'{e}e Kirst}, \bibinfo{person}{Qi Wang}, {and} \bibinfo{person}{Yun Huang}.} \bibinfo{year}{2023}\natexlab{}.
\newblock \showarticletitle{Exploring Think-aloud Method with Deaf and Hard of Hearing College Students}. In \bibinfo{booktitle}{\emph{Proceedings of the 2023 ACM Designing Interactive Systems Conference}} (Pittsburgh, PA, USA) \emph{(\bibinfo{series}{DIS '23})}. \bibinfo{publisher}{Association for Computing Machinery}, \bibinfo{address}{New York, NY, USA}, \bibinfo{pages}{1757–1772}.
\newblock
\showISBNx{9781450398930}
\urldef\tempurl%
\url{https://doi.org/10.1145/3563657.3595980}
\showDOI{\tempurl}


\bibitem[Chen et~al\mbox{.}(2024c)]%
        {chen2024mixed}
\bibfield{author}{\bibinfo{person}{Si Chen}, \bibinfo{person}{James Waller}, \bibinfo{person}{Matthew Seita}, \bibinfo{person}{Christian Vogler}, \bibinfo{person}{Raja Kushalnagar}, {and} \bibinfo{person}{Qi Wang}.} \bibinfo{year}{2024}\natexlab{c}.
\newblock \showarticletitle{Towards Co-Creating Access and Inclusion: A Group Autoethnography on a Hearing Individual's Journey Towards Effective Communication in Mixed-Hearing Ability Higher Education Settings}. In \bibinfo{booktitle}{\emph{Proceedings of the CHI Conference on Human Factors in Computing Systems}} (Honolulu, HI, USA) \emph{(\bibinfo{series}{CHI '24})}. \bibinfo{publisher}{Association for Computing Machinery}, \bibinfo{address}{New York, NY, USA}, Article \bibinfo{articleno}{55}, \bibinfo{numpages}{14}~pages.
\newblock
\showISBNx{9798400703300}
\urldef\tempurl%
\url{https://doi.org/10.1145/3613904.3642017}
\showDOI{\tempurl}


\bibitem[Choi et~al\mbox{.}(2024)]%
        {choi2024vivid}
\bibfield{author}{\bibinfo{person}{Seulgi Choi}, \bibinfo{person}{Hyewon Lee}, \bibinfo{person}{Yoonjoo Lee}, {and} \bibinfo{person}{Juho Kim}.} \bibinfo{year}{2024}\natexlab{}.
\newblock \showarticletitle{VIVID: Human-AI Collaborative Authoring of Vicarious Dialogues from Lecture Videos}. In \bibinfo{booktitle}{\emph{Proceedings of the CHI Conference on Human Factors in Computing Systems}}. \bibinfo{pages}{1--26}.
\newblock


\bibitem[Cotton(1988)]%
        {cotton1988classroom}
\bibfield{author}{\bibinfo{person}{Kathleen Cotton}.} \bibinfo{year}{1988}\natexlab{}.
\newblock \showarticletitle{Classroom questioning}.
\newblock \bibinfo{journal}{\emph{School improvement research series}}  \bibinfo{volume}{5} (\bibinfo{year}{1988}), \bibinfo{pages}{1--22}.
\newblock


\bibitem[Coy et~al\mbox{.}(2024)]%
        {Coy2024education}
\bibfield{author}{\bibinfo{person}{André Coy}, \bibinfo{person}{Phaedra~S. Mohammed}, {and} \bibinfo{person}{Paulson Skerrit}.} \bibinfo{year}{2024}\natexlab{}.
\newblock \showarticletitle{Inclusive Deaf Education Enabled by Artificial Intelligence: The Path to a Solution}.
\newblock \bibinfo{journal}{\emph{International Journal of Artificial Intelligence in Education}} (\bibinfo{year}{2024}).
\newblock
\showISSN{1560-4306}
\urldef\tempurl%
\url{https://doi.org/10.1007/s40593-024-00419-9}
\showDOI{\tempurl}


\bibitem[Cui and Sachan(2023)]%
        {cui2023adaptive}
\bibfield{author}{\bibinfo{person}{Peng Cui} {and} \bibinfo{person}{Mrinmaya Sachan}.} \bibinfo{year}{2023}\natexlab{}.
\newblock \showarticletitle{Adaptive and personalized exercise generation for online language learning}.
\newblock \bibinfo{journal}{\emph{arXiv preprint arXiv:2306.02457}} (\bibinfo{year}{2023}).
\newblock


\bibitem[Debevc and Živa Peljhan(2004)]%
        {Debevc2004roleofvideo}
\bibfield{author}{\bibinfo{person}{Matjaž Debevc} {and} \bibinfo{person}{Živa Peljhan}.} \bibinfo{year}{2004}\natexlab{}.
\newblock \showarticletitle{The role of video technology in on-line lectures for the deaf}.
\newblock \bibinfo{journal}{\emph{Disability and Rehabilitation}} \bibinfo{volume}{26}, \bibinfo{number}{17} (\bibinfo{year}{2004}), \bibinfo{pages}{1048--1059}.
\newblock
\urldef\tempurl%
\url{https://doi.org/10.1080/09638280410001702441}
\showDOI{\tempurl}
\showeprint{https://doi.org/10.1080/09638280410001702441}
\newblock
\shownote{PMID: 15371041}.


\bibitem[Desai et~al\mbox{.}(2024)]%
        {desai2024systemic}
\bibfield{author}{\bibinfo{person}{Aashaka Desai}, \bibinfo{person}{Maartje De~Meulder}, \bibinfo{person}{Julie~A. Hochgesang}, \bibinfo{person}{Annemarie Kocab}, {and} \bibinfo{person}{Alex~X. Lu}.} \bibinfo{year}{2024}\natexlab{}.
\newblock \showarticletitle{Systemic Biases in Sign Language {AI} Research: A Deaf-Led Call to Reevaluate Research Agendas}. In \bibinfo{booktitle}{\emph{Proceedings of the LREC-COLING 2024 11th Workshop on the Representation and Processing of Sign Languages: Evaluation of Sign Language Resources}}, \bibfield{editor}{\bibinfo{person}{Eleni Efthimiou}, \bibinfo{person}{Stavroula-Evita Fotinea}, \bibinfo{person}{Thomas Hanke}, \bibinfo{person}{Julie~A. Hochgesang}, \bibinfo{person}{Johanna Mesch}, {and} \bibinfo{person}{Marc Schulder}} (Eds.). \bibinfo{publisher}{ELRA and ICCL}, \bibinfo{address}{Torino, Italia}, \bibinfo{pages}{54--65}.
\newblock
\urldef\tempurl%
\url{https://aclanthology.org/2024.signlang-1.6}
\showURL{%
\tempurl}


\bibitem[Draxler et~al\mbox{.}(2023)]%
        {draxler2023relevance}
\bibfield{author}{\bibinfo{person}{Fiona Draxler}, \bibinfo{person}{Albrecht Schmidt}, {and} \bibinfo{person}{Lewis~L Chuang}.} \bibinfo{year}{2023}\natexlab{}.
\newblock \showarticletitle{Relevance, Effort, and Perceived Quality: Language Learners’ Experiences with AI-Generated Contextually Personalized Learning Material}. In \bibinfo{booktitle}{\emph{Proceedings of the 2023 ACM Designing Interactive Systems Conference}}. \bibinfo{pages}{2249--2262}.
\newblock


\bibitem[Drigas et~al\mbox{.}(2005)]%
        {Drigas2005elearning}
\bibfield{author}{\bibinfo{person}{A.S. Drigas}, \bibinfo{person}{D. Kouremenos}, \bibinfo{person}{S. Kouremenos}, {and} \bibinfo{person}{J. Vrettaros}.} \bibinfo{year}{2005}\natexlab{}.
\newblock \showarticletitle{An e-learning system for the deaf people}. In \bibinfo{booktitle}{\emph{2005 6th International Conference on Information Technology Based Higher Education and Training}}. \bibinfo{pages}{T2C/17--T2C/21}.
\newblock
\urldef\tempurl%
\url{https://doi.org/10.1109/ITHET.2005.1560236}
\showDOI{\tempurl}


\bibitem[Fung et~al\mbox{.}(2024)]%
        {fung2024automatic}
\bibfield{author}{\bibinfo{person}{Sze Ching~Evelyn Fung}, \bibinfo{person}{Man~Fai Wong}, {and} \bibinfo{person}{Chee~Wei Tan}.} \bibinfo{year}{2024}\natexlab{}.
\newblock \showarticletitle{Automatic Feedback Generation on K-12 Students' Data Science Education by Prompting Cloud-based Large Language Models}. In \bibinfo{booktitle}{\emph{Proceedings of the Eleventh ACM Conference on Learning@ Scale}}. \bibinfo{pages}{255--258}.
\newblock


\bibitem[Guardino and Cannon(2016)]%
        {guardino2016deafness}
\bibfield{author}{\bibinfo{person}{Caroline Guardino} {and} \bibinfo{person}{Joanna~E Cannon}.} \bibinfo{year}{2016}\natexlab{}.
\newblock \showarticletitle{Deafness and diversity: Reflections and directions}.
\newblock \bibinfo{journal}{\emph{American Annals of the Deaf}} \bibinfo{volume}{161}, \bibinfo{number}{1} (\bibinfo{year}{2016}), \bibinfo{pages}{104--112}.
\newblock


\bibitem[Holman et~al\mbox{.}(2024)]%
        {holman2024navigating}
\bibfield{author}{\bibinfo{person}{Kenneth Holman}, \bibinfo{person}{Matthew Marino}, \bibinfo{person}{Eleazar Vasquez}, \bibinfo{person}{Michelle Taub}, \bibinfo{person}{Jessica Hunt}, {and} \bibinfo{person}{Yacine Tazi}.} \bibinfo{year}{2024}\natexlab{}.
\newblock \showarticletitle{Navigating AI-Powered Personalized Learning in Special Education: A Guide for Preservice Teacher Faculty}.
\newblock \bibinfo{journal}{\emph{Journal of Special Education Preparation}} \bibinfo{volume}{4}, \bibinfo{number}{2} (\bibinfo{year}{2024}), \bibinfo{pages}{90--95}.
\newblock


\bibitem[Hutt and Hieb(2024)]%
        {hutt2024scaling}
\bibfield{author}{\bibinfo{person}{Stephen Hutt} {and} \bibinfo{person}{Grayson Hieb}.} \bibinfo{year}{2024}\natexlab{}.
\newblock \showarticletitle{Scaling Up Mastery Learning with Generative AI: Exploring How Generative AI Can Assist in the Generation and Evaluation of Mastery Quiz Questions}. In \bibinfo{booktitle}{\emph{Proceedings of the Eleventh ACM Conference on Learning@ Scale}}. \bibinfo{pages}{310--314}.
\newblock


\bibitem[Kafle et~al\mbox{.}(2020)]%
        {Kafle2019fairness}
\bibfield{author}{\bibinfo{person}{Sushant Kafle}, \bibinfo{person}{Abraham Glasser}, \bibinfo{person}{Sedeeq Al-khazraji}, \bibinfo{person}{Larwan Berke}, \bibinfo{person}{Matthew Seita}, {and} \bibinfo{person}{Matt Huenerfauth}.} \bibinfo{year}{2020}\natexlab{}.
\newblock \showarticletitle{Artificial intelligence fairness in the context of accessibility research on intelligent systems for people who are deaf or hard of hearing}.
\newblock \bibinfo{journal}{\emph{SIGACCESS Access. Comput.}} \bibinfo{number}{125}, Article \bibinfo{articleno}{4} (\bibinfo{date}{mar} \bibinfo{year}{2020}), \bibinfo{numpages}{1}~pages.
\newblock
\showISSN{1558-2337}
\urldef\tempurl%
\url{https://doi.org/10.1145/3386296.3386300}
\showDOI{\tempurl}


\bibitem[Karpicke and Blunt(2011)]%
        {karpicke2011retrieval}
\bibfield{author}{\bibinfo{person}{Jeffrey~D Karpicke} {and} \bibinfo{person}{Janell~R Blunt}.} \bibinfo{year}{2011}\natexlab{}.
\newblock \showarticletitle{Retrieval practice produces more learning than elaborative studying with concept mapping}.
\newblock \bibinfo{journal}{\emph{Science}} \bibinfo{volume}{331}, \bibinfo{number}{6018} (\bibinfo{year}{2011}), \bibinfo{pages}{772--775}.
\newblock


\bibitem[Kasneci et~al\mbox{.}(2023)]%
        {kasneci2023chatgpt}
\bibfield{author}{\bibinfo{person}{Enkelejda Kasneci}, \bibinfo{person}{Kathrin Sessler}, \bibinfo{person}{Stefan Küchemann}, \bibinfo{person}{Maria Bannert}, \bibinfo{person}{Daryna Dementieva}, \bibinfo{person}{Frank Fischer}, \bibinfo{person}{Urs Gasser}, \bibinfo{person}{Georg Groh}, \bibinfo{person}{Stephan Günnemann}, \bibinfo{person}{Eyke Hüllermeier}, \bibinfo{person}{Stephan Krusche}, \bibinfo{person}{Gitta Kutyniok}, \bibinfo{person}{Tilman Michaeli}, \bibinfo{person}{Claudia Nerdel}, \bibinfo{person}{Jürgen Pfeffer}, \bibinfo{person}{Oleksandra Poquet}, \bibinfo{person}{Michael Sailer}, \bibinfo{person}{Albrecht Schmidt}, \bibinfo{person}{Tina Seidel}, \bibinfo{person}{Matthias Stadler}, \bibinfo{person}{Jochen Weller}, \bibinfo{person}{Jochen Kuhn}, {and} \bibinfo{person}{Gjergji Kasneci}.} \bibinfo{year}{2023}\natexlab{}.
\newblock \showarticletitle{ChatGPT for good? On opportunities and challenges of large language models for education}.
\newblock \bibinfo{journal}{\emph{Learning and Individual Differences}}  \bibinfo{volume}{103} (\bibinfo{year}{2023}), \bibinfo{pages}{102274}.
\newblock
\showISSN{1041-6080}
\urldef\tempurl%
\url{https://doi.org/10.1016/j.lindif.2023.102274}
\showDOI{\tempurl}


\bibitem[Khasawneh(2023)]%
        {Khasawneh2023videomedia}
\bibfield{author}{\bibinfo{person}{Mohamad Ahmad~Saleem Khasawneh}.} \bibinfo{year}{2023}\natexlab{}.
\newblock \showarticletitle{The use of video as media in distance learning for deaf students}.
\newblock \bibinfo{journal}{\emph{Contemporary Educational Technology}} \bibinfo{volume}{15}, \bibinfo{number}{2} (\bibinfo{year}{2023}), \bibinfo{pages}{ep418}.
\newblock
\urldef\tempurl%
\url{https://doi.org/10.30935/cedtech/13012}
\showDOI{\tempurl}


\bibitem[Kurdi et~al\mbox{.}(2020)]%
        {kurdi2020systematic}
\bibfield{author}{\bibinfo{person}{Ghader Kurdi}, \bibinfo{person}{Jared Leo}, \bibinfo{person}{Bijan Parsia}, \bibinfo{person}{Uli Sattler}, {and} \bibinfo{person}{Salam Al-Emari}.} \bibinfo{year}{2020}\natexlab{}.
\newblock \showarticletitle{A systematic review of automatic question generation for educational purposes}.
\newblock \bibinfo{journal}{\emph{International Journal of Artificial Intelligence in Education}}  \bibinfo{volume}{30} (\bibinfo{year}{2020}), \bibinfo{pages}{121--204}.
\newblock


\bibitem[Kushalnagar et~al\mbox{.}(2012)]%
        {kushalnagar2012readability}
\bibfield{author}{\bibinfo{person}{Raja~S. Kushalnagar}, \bibinfo{person}{Walter~S. Lasecki}, {and} \bibinfo{person}{Jeffrey~P. Bigham}.} \bibinfo{year}{2012}\natexlab{}.
\newblock \showarticletitle{A readability evaluation of real-time crowd captions in the classroom}. In \bibinfo{booktitle}{\emph{Proceedings of the 14th International ACM SIGACCESS Conference on Computers and Accessibility}} (Boulder, Colorado, USA) \emph{(\bibinfo{series}{ASSETS '12})}. \bibinfo{publisher}{Association for Computing Machinery}, \bibinfo{address}{New York, NY, USA}, \bibinfo{pages}{71–78}.
\newblock
\showISBNx{9781450313216}
\urldef\tempurl%
\url{https://doi.org/10.1145/2384916.2384930}
\showDOI{\tempurl}


\bibitem[Lasecki et~al\mbox{.}(2014)]%
        {lasecki2014caption}
\bibfield{author}{\bibinfo{person}{Walter~S. Lasecki}, \bibinfo{person}{Raja Kushalnagar}, {and} \bibinfo{person}{Jeffrey~P. Bigham}.} \bibinfo{year}{2014}\natexlab{}.
\newblock \showarticletitle{Helping students keep up with real-time captions by pausing and highlighting}. In \bibinfo{booktitle}{\emph{Proceedings of the 11th Web for All Conference}} (Seoul, Korea) \emph{(\bibinfo{series}{W4A '14})}. \bibinfo{publisher}{Association for Computing Machinery}, \bibinfo{address}{New York, NY, USA}, Article \bibinfo{articleno}{39}, \bibinfo{numpages}{8}~pages.
\newblock
\showISBNx{9781450326513}
\urldef\tempurl%
\url{https://doi.org/10.1145/2596695.2596701}
\showDOI{\tempurl}


\bibitem[Lindberg et~al\mbox{.}(2013)]%
        {lindberg2013generating}
\bibfield{author}{\bibinfo{person}{David Lindberg}, \bibinfo{person}{Fred Popowich}, \bibinfo{person}{John Nesbit}, {and} \bibinfo{person}{Phil Winne}.} \bibinfo{year}{2013}\natexlab{}.
\newblock \showarticletitle{Generating natural language questions to support learning on-line}. In \bibinfo{booktitle}{\emph{Proceedings of the 14th European workshop on natural language generation}}. \bibinfo{pages}{105--114}.
\newblock


\bibitem[Lu et~al\mbox{.}(2023)]%
        {lu2023readingquizmaker}
\bibfield{author}{\bibinfo{person}{Xinyi Lu}, \bibinfo{person}{Simin Fan}, \bibinfo{person}{Jessica Houghton}, \bibinfo{person}{Lu Wang}, {and} \bibinfo{person}{Xu Wang}.} \bibinfo{year}{2023}\natexlab{}.
\newblock \showarticletitle{ReadingQuizMaker: a human-NLP collaborative system that supports instructors to design high-quality reading quiz questions}. In \bibinfo{booktitle}{\emph{Proceedings of the 2023 CHI Conference on Human Factors in Computing Systems}}. \bibinfo{pages}{1--18}.
\newblock


\bibitem[Mack et~al\mbox{.}(2024)]%
        {mack2024wheelchair}
\bibfield{author}{\bibinfo{person}{Avery Mack}, \bibinfo{person}{Rida Qadri}, \bibinfo{person}{Remi Denton}, \bibinfo{person}{Shaun~K Kane}, {and} \bibinfo{person}{Cynthia~L Bennett}.} \bibinfo{year}{2024}\natexlab{}.
\newblock \showarticletitle{“They only care to show us the wheelchair”: disability representation in text-to-image AI models}.
\newblock
\urldef\tempurl%
\url{https://drive.google.com/file/d/1Fys0pKsAFqN9zY8LcxxL2lTUmn01xU73/view?usp=sharing}
\showURL{%
\tempurl}


\bibitem[Mayer(2014)]%
        {mayer20143}
\bibfield{author}{\bibinfo{person}{Richard~E Mayer}.} \bibinfo{year}{2014}\natexlab{}.
\newblock \showarticletitle{3 Cognitive Theory of Multimedia Learning}.
\newblock \bibinfo{journal}{\emph{The Cambridge Handbook of Multimedia Learning}} (\bibinfo{year}{2014}), \bibinfo{pages}{43}.
\newblock


\bibitem[Mooney and Lashewicz(2015)]%
        {mooney2015love}
\bibfield{author}{\bibinfo{person}{Laura~Rae Mooney} {and} \bibinfo{person}{Bonnie Lashewicz}.} \bibinfo{year}{2015}\natexlab{}.
\newblock \showarticletitle{For the love of the child: Bestowing value amidst inconsistent inclusive education beliefs and practices for one student with severe disabilities}.
\newblock \bibinfo{journal}{\emph{Canadian Journal of Education/Revue canadienne de l'{\'e}ducation}} \bibinfo{volume}{38}, \bibinfo{number}{4} (\bibinfo{year}{2015}), \bibinfo{pages}{1--28}.
\newblock


\bibitem[Nagata(2019)]%
        {nagata2019toward}
\bibfield{author}{\bibinfo{person}{Ryo Nagata}.} \bibinfo{year}{2019}\natexlab{}.
\newblock \showarticletitle{Toward a task of feedback comment generation for writing learning}. In \bibinfo{booktitle}{\emph{Proceedings of the 2019 Conference on Empirical Methods in Natural Language Processing and the 9th International Joint Conference on Natural Language Processing (EMNLP-IJCNLP)}}. \bibinfo{pages}{3206--3215}.
\newblock


\bibitem[Nagata et~al\mbox{.}(2021)]%
        {nagata2021shared}
\bibfield{author}{\bibinfo{person}{Ryo Nagata}, \bibinfo{person}{Masato Hagiwara}, \bibinfo{person}{Kazuaki Hanawa}, \bibinfo{person}{Masato Mita}, \bibinfo{person}{Artem Chernodub}, {and} \bibinfo{person}{Olena Nahorna}.} \bibinfo{year}{2021}\natexlab{}.
\newblock \showarticletitle{Shared task on feedback comment generation for language learners}. In \bibinfo{booktitle}{\emph{Proceedings of the 14th International Conference on Natural Language Generation}}. \bibinfo{pages}{320--324}.
\newblock


\bibitem[Niebuhr et~al\mbox{.}(2014)]%
        {niebuhr2014online}
\bibfield{author}{\bibinfo{person}{Virginia Niebuhr}, \bibinfo{person}{Bruce Niebuhr}, \bibinfo{person}{Julie Trumble}, {and} \bibinfo{person}{Mary~Jo Urbani}.} \bibinfo{year}{2014}\natexlab{}.
\newblock \showarticletitle{Online faculty development for creating E-learning materials}.
\newblock \bibinfo{journal}{\emph{Education for health}} \bibinfo{volume}{27}, \bibinfo{number}{3} (\bibinfo{year}{2014}), \bibinfo{pages}{255--261}.
\newblock


\bibitem[Nomura et~al\mbox{.}(2006)]%
        {nomura2006measurement}
\bibfield{author}{\bibinfo{person}{Tatsuya Nomura}, \bibinfo{person}{Tomohiro Suzuki}, \bibinfo{person}{Takayuki Kanda}, {and} \bibinfo{person}{Kensuke Kato}.} \bibinfo{year}{2006}\natexlab{}.
\newblock \showarticletitle{Measurement of negative attitudes toward robots}.
\newblock \bibinfo{journal}{\emph{Interaction Studies. Social Behaviour and Communication in Biological and Artificial Systems}} \bibinfo{volume}{7}, \bibinfo{number}{3} (\bibinfo{year}{2006}), \bibinfo{pages}{437--454}.
\newblock


\bibitem[Paudyal et~al\mbox{.}(2019)]%
        {paudyal2018daveeVR}
\bibfield{author}{\bibinfo{person}{Prajwal Paudyal}, \bibinfo{person}{Ayan Banerjee}, \bibinfo{person}{Yijian Hu}, {and} \bibinfo{person}{Sandeep Gupta}.} \bibinfo{year}{2019}\natexlab{}.
\newblock \showarticletitle{DAVEE: A Deaf Accessible Virtual Environment for Education}. In \bibinfo{booktitle}{\emph{Proceedings of the 2019 Conference on Creativity and Cognition}} (San Diego, CA, USA) \emph{(\bibinfo{series}{C\&C '19})}. \bibinfo{publisher}{Association for Computing Machinery}, \bibinfo{address}{New York, NY, USA}, \bibinfo{pages}{522–526}.
\newblock
\showISBNx{9781450359177}
\urldef\tempurl%
\url{https://doi.org/10.1145/3325480.3326546}
\showDOI{\tempurl}


\bibitem[Pratiwi et~al\mbox{.}(2019)]%
        {Pratiwi_2019}
\bibfield{author}{\bibinfo{person}{AS Pratiwi}, \bibinfo{person}{AT Lestari}, \bibinfo{person}{B Hendrawan}, \bibinfo{person}{MF Nugraha}, \bibinfo{person}{M. Nurfitriani}, \bibinfo{person}{M Nurkamilah}, \bibinfo{person}{Mujiarto}, \bibinfo{person}{Tadkiroatun Musfiroh}, \bibinfo{person}{F Nugraha}, {and} \bibinfo{person}{Wan~Ridwan H}.} \bibinfo{year}{2019}\natexlab{}.
\newblock \showarticletitle{Digital Video Based Rampak Kendang Learning Media for Deaf Students}.
\newblock \bibinfo{journal}{\emph{Journal of Physics: Conference Series}} \bibinfo{volume}{1179}, \bibinfo{number}{1} (\bibinfo{date}{jul} \bibinfo{year}{2019}), \bibinfo{pages}{012040}.
\newblock
\urldef\tempurl%
\url{https://doi.org/10.1088/1742-6596/1179/1/012040}
\showDOI{\tempurl}


\bibitem[Reed et~al\mbox{.}(2008)]%
        {Reed2008academicstatus}
\bibfield{author}{\bibinfo{person}{Susanne Reed}, \bibinfo{person}{Shirin~D. Antia}, {and} \bibinfo{person}{Kathryn~H. Kreimeyer}.} \bibinfo{year}{2008}\natexlab{}.
\newblock \showarticletitle{{Academic Status of Deaf and Hard-of-Hearing Students in Public Schools: Student, Home, and Service Facilitators and Detractors}}.
\newblock \bibinfo{journal}{\emph{The Journal of Deaf Studies and Deaf Education}} \bibinfo{volume}{13}, \bibinfo{number}{4} (\bibinfo{date}{03} \bibinfo{year}{2008}), \bibinfo{pages}{485--502}.
\newblock
\showISSN{1081-4159}
\urldef\tempurl%
\url{https://doi.org/10.1093/deafed/enn006}
\showDOI{\tempurl}
\showeprint{https://academic.oup.com/jdsde/article-pdf/13/4/485/1338266/enn006.pdf}


\bibitem[Scardamalia et~al\mbox{.}(2019)]%
        {scardamalia2019consistently}
\bibfield{author}{\bibinfo{person}{Kristin Scardamalia}, \bibinfo{person}{Keisha~L Bentley-Edwards}, {and} \bibinfo{person}{Kairys Grasty}.} \bibinfo{year}{2019}\natexlab{}.
\newblock \showarticletitle{Consistently inconsistent: An examination of the variability in the identification of emotional disturbance}.
\newblock \bibinfo{journal}{\emph{Psychology in the Schools}} \bibinfo{volume}{56}, \bibinfo{number}{4} (\bibinfo{year}{2019}), \bibinfo{pages}{569--581}.
\newblock


\bibitem[Scott et~al\mbox{.}(2021)]%
        {Scott2021callfordiversity}
\bibfield{author}{\bibinfo{person}{Jessica~A. Scott}, \bibinfo{person}{Joanna~E. Dostal}, {and} \bibinfo{person}{Katharine Lane-Outlaw}.} \bibinfo{year}{2021}\natexlab{}.
\newblock \showarticletitle{A Call for a Diversity of Perspectives in Deaf Education Research: A Response to Mayer and Trezek (2020)}.
\newblock \bibinfo{journal}{\emph{American Annals of the Deaf}} \bibinfo{volume}{166}, \bibinfo{number}{1} (\bibinfo{year}{2021}), \bibinfo{pages}{49--61}.
\newblock
\urldef\tempurl%
\url{https://www.jstor.org/stable/27087035}
\showURL{%
\tempurl}
\newblock
\shownote{Accessed 31 Aug. 2024}.


\bibitem[Sedláčková and HOLMQVIST(2020)]%
        {sedlackova2020textbook}
\bibfield{author}{\bibinfo{person}{Jitka Sedláčková} {and} \bibinfo{person}{Kenneth HOLMQVIST}.} \bibinfo{year}{2020}\natexlab{}.
\newblock \showarticletitle{Deaf learners' processing of English textbook material : An eye-tracking study}.
\newblock


\bibitem[Shelton and Parlin(2016)]%
        {shelton2016math}
\bibfield{author}{\bibinfo{person}{Brett Shelton} {and} \bibinfo{person}{Mary Parlin}.} \bibinfo{year}{2016}\natexlab{}.
\newblock \showarticletitle{Teaching Math to Deaf/Hard-of-Hearing (DHH) Children Using Mobile Games:}.
\newblock \bibinfo{journal}{\emph{International Journal of Mobile and Blended Learning}}  \bibinfo{volume}{8} (\bibinfo{date}{01} \bibinfo{year}{2016}), \bibinfo{pages}{1--17}.
\newblock
\urldef\tempurl%
\url{https://doi.org/10.4018/IJMBL.2016010101}
\showDOI{\tempurl}


\bibitem[Shin et~al\mbox{.}(2018)]%
        {shin2018understanding}
\bibfield{author}{\bibinfo{person}{Hyungyu Shin}, \bibinfo{person}{Eun-Young Ko}, \bibinfo{person}{Joseph~Jay Williams}, {and} \bibinfo{person}{Juho Kim}.} \bibinfo{year}{2018}\natexlab{}.
\newblock \showarticletitle{Understanding the effect of in-video prompting on learners and instructors}. In \bibinfo{booktitle}{\emph{Proceedings of the 2018 CHI conference on human factors in computing systems}}. \bibinfo{pages}{1--12}.
\newblock


\bibitem[Stowe et~al\mbox{.}(2022)]%
        {stowe2022controlled}
\bibfield{author}{\bibinfo{person}{Kevin Stowe}, \bibinfo{person}{Debanjan Ghosh}, {and} \bibinfo{person}{Mengxuan Zhao}.} \bibinfo{year}{2022}\natexlab{}.
\newblock \showarticletitle{Controlled language generation for language learning items}.
\newblock \bibinfo{journal}{\emph{arXiv preprint arXiv:2211.15731}} (\bibinfo{year}{2022}).
\newblock


\bibitem[Team et~al\mbox{.}(2023)]%
        {team2023gemini}
\bibfield{author}{\bibinfo{person}{Gemini Team}, \bibinfo{person}{Rohan Anil}, \bibinfo{person}{Sebastian Borgeaud}, \bibinfo{person}{Yonghui Wu}, \bibinfo{person}{Jean-Baptiste Alayrac}, \bibinfo{person}{Jiahui Yu}, \bibinfo{person}{Radu Soricut}, \bibinfo{person}{Johan Schalkwyk}, \bibinfo{person}{Andrew~M Dai}, \bibinfo{person}{Anja Hauth}, {et~al\mbox{.}}} \bibinfo{year}{2023}\natexlab{}.
\newblock \showarticletitle{Gemini: a family of highly capable multimodal models}.
\newblock \bibinfo{journal}{\emph{arXiv preprint arXiv:2312.11805}} (\bibinfo{year}{2023}).
\newblock


\bibitem[Touvron et~al\mbox{.}(2023)]%
        {touvron2023llama}
\bibfield{author}{\bibinfo{person}{Hugo Touvron}, \bibinfo{person}{Thibaut Lavril}, \bibinfo{person}{Gautier Izacard}, \bibinfo{person}{Xavier Martinet}, \bibinfo{person}{Marie-Anne Lachaux}, \bibinfo{person}{Timoth{\'e}e Lacroix}, \bibinfo{person}{Baptiste Rozi{\`e}re}, \bibinfo{person}{Naman Goyal}, \bibinfo{person}{Eric Hambro}, \bibinfo{person}{Faisal Azhar}, {et~al\mbox{.}}} \bibinfo{year}{2023}\natexlab{}.
\newblock \showarticletitle{Llama: Open and efficient foundation language models}.
\newblock \bibinfo{journal}{\emph{arXiv preprint arXiv:2302.13971}} (\bibinfo{year}{2023}).
\newblock


\bibitem[van~der Meij and Bockmann(2021)]%
        {van2021effects}
\bibfield{author}{\bibinfo{person}{Hans van~der Meij} {and} \bibinfo{person}{Linn Bockmann}.} \bibinfo{year}{2021}\natexlab{}.
\newblock \showarticletitle{Effects of embedded questions in recorded lectures}.
\newblock \bibinfo{journal}{\emph{Journal of computing in higher education}} \bibinfo{volume}{33}, \bibinfo{number}{1} (\bibinfo{year}{2021}), \bibinfo{pages}{235--254}.
\newblock


\bibitem[Wang and Piper(2018)]%
        {wang2018a11y}
\bibfield{author}{\bibinfo{person}{Emily~Q. Wang} {and} \bibinfo{person}{Anne~Marie Piper}.} \bibinfo{year}{2018}\natexlab{}.
\newblock \showarticletitle{Accessibility in Action: Co-Located Collaboration among Deaf and Hearing Professionals}.
\newblock \bibinfo{journal}{\emph{Proc. ACM Hum.-Comput. Interact.}} \bibinfo{volume}{2}, \bibinfo{number}{CSCW}, Article \bibinfo{articleno}{180} (\bibinfo{date}{nov} \bibinfo{year}{2018}), \bibinfo{numpages}{25}~pages.
\newblock
\urldef\tempurl%
\url{https://doi.org/10.1145/3274449}
\showDOI{\tempurl}


\bibitem[Wang et~al\mbox{.}(2021)]%
        {wang2021seeing}
\bibfield{author}{\bibinfo{person}{Xu Wang}, \bibinfo{person}{Carolyn Rose}, {and} \bibinfo{person}{Ken Koedinger}.} \bibinfo{year}{2021}\natexlab{}.
\newblock \showarticletitle{Seeing beyond expert blind spots: Online learning design for scale and quality}. In \bibinfo{booktitle}{\emph{Proceedings of the 2021 CHI Conference on Human Factors in Computing Systems}}. \bibinfo{pages}{1--14}.
\newblock


\bibitem[Wang et~al\mbox{.}(2019)]%
        {wang2019upgrade}
\bibfield{author}{\bibinfo{person}{Xu Wang}, \bibinfo{person}{Srinivasa~Teja Talluri}, \bibinfo{person}{Carolyn Rose}, {and} \bibinfo{person}{Kenneth Koedinger}.} \bibinfo{year}{2019}\natexlab{}.
\newblock \showarticletitle{UpGrade: Sourcing student open-ended solutions to create scalable learning opportunities}. In \bibinfo{booktitle}{\emph{Proceedings of the Sixth (2019) ACM Conference on Learning@ Scale}}. \bibinfo{pages}{1--10}.
\newblock


\bibitem[Wu et~al\mbox{.}(2022)]%
        {wu2022survey}
\bibfield{author}{\bibinfo{person}{Xingjiao Wu}, \bibinfo{person}{Luwei Xiao}, \bibinfo{person}{Yixuan Sun}, \bibinfo{person}{Junhang Zhang}, \bibinfo{person}{Tianlong Ma}, {and} \bibinfo{person}{Liang He}.} \bibinfo{year}{2022}\natexlab{}.
\newblock \showarticletitle{A survey of human-in-the-loop for machine learning}.
\newblock \bibinfo{journal}{\emph{Future Generation Computer Systems}}  \bibinfo{volume}{135} (\bibinfo{year}{2022}), \bibinfo{pages}{364--381}.
\newblock


\bibitem[Yang et~al\mbox{.}(2024)]%
        {yang2024aqua}
\bibfield{author}{\bibinfo{person}{Saelyne Yang}, \bibinfo{person}{Jo Vermeulen}, \bibinfo{person}{George Fitzmaurice}, {and} \bibinfo{person}{Justin Matejka}.} \bibinfo{year}{2024}\natexlab{}.
\newblock \showarticletitle{AQuA: Automated Question-Answering in Software Tutorial Videos with Visual Anchors}. In \bibinfo{booktitle}{\emph{Proceedings of the CHI Conference on Human Factors in Computing Systems}}. \bibinfo{pages}{1--19}.
\newblock


\bibitem[Zamfirescu-Pereira et~al\mbox{.}(2023)]%
        {zamfirescu2023johnny}
\bibfield{author}{\bibinfo{person}{JD Zamfirescu-Pereira}, \bibinfo{person}{Richmond~Y Wong}, \bibinfo{person}{Bjoern Hartmann}, {and} \bibinfo{person}{Qian Yang}.} \bibinfo{year}{2023}\natexlab{}.
\newblock \showarticletitle{Why Johnny can’t prompt: how non-AI experts try (and fail) to design LLM prompts}. In \bibinfo{booktitle}{\emph{Proceedings of the 2023 CHI Conference on Human Factors in Computing Systems}}. \bibinfo{pages}{1--21}.
\newblock


\bibitem[Zirzow(2015)]%
        {zirzow2015avatar}
\bibfield{author}{\bibinfo{person}{Nichole~K. Zirzow}.} \bibinfo{year}{2015}\natexlab{}.
\newblock \showarticletitle{Signing Avatars: Using Virtual Reality to Support Students with Hearing Loss}.
\newblock \bibinfo{journal}{\emph{Rural Special Education Quarterly}} \bibinfo{volume}{34}, \bibinfo{number}{3} (\bibinfo{year}{2015}), \bibinfo{pages}{33--36}.
\newblock
\urldef\tempurl%
\url{https://doi.org/10.1177/875687051503400307}
\showDOI{\tempurl}
\showeprint{https://doi.org/10.1177/875687051503400307}


\end{thebibliography}

\newpage
\section{Appendices}

\subsection{Prototype Architecture}

The full-stack web application is implemented using Flask framework\footnote{https://flask.palletsprojects.com/en/3.0.x/} as the web server, featuring the following key functionalities (Fig. 4):
\begin{enumerate}
\item Chatbot Integration: The OpenAI GPT-4 API powers our chatbot.
\item Question Selection: When participants indicate their category preferences, the prototype automatically selects questions based on their input. The questions are stored in JSON format with fields including "\texttt{timestamp}", "\texttt{question}", "\texttt{answers}", "\texttt{transcript\_timestamp\_start}", and "\texttt{transcript\_reference}." The "\texttt{timestamp}" field specifies when the question should appear during video playback, while the "\texttt{transcript\_reference}" allows participants to rewind the video if needed.
\item Video Playback and Question Pop-Up: JavaScript listeners monitor the video timestamps. When a match occurs, the controller triggers the pop-up questions with all registered information (Fig. \ref{fig:system_screens}).
\item Post-Video Rating: After the video, participants rate the questions they answered. This step is for user experience collection which is explained in section \ref{rate}.

All interaction data and participant choices are recorded in an Azure database. This comprehensive prototype is deployed on Microsoft Azure Cloud Service\footnote{https://azure.microsoft.com/en-us/}.
\end{enumerate}

\begin{figure}[ht]
    \centering  
    \includegraphics[width=0.4\textwidth]{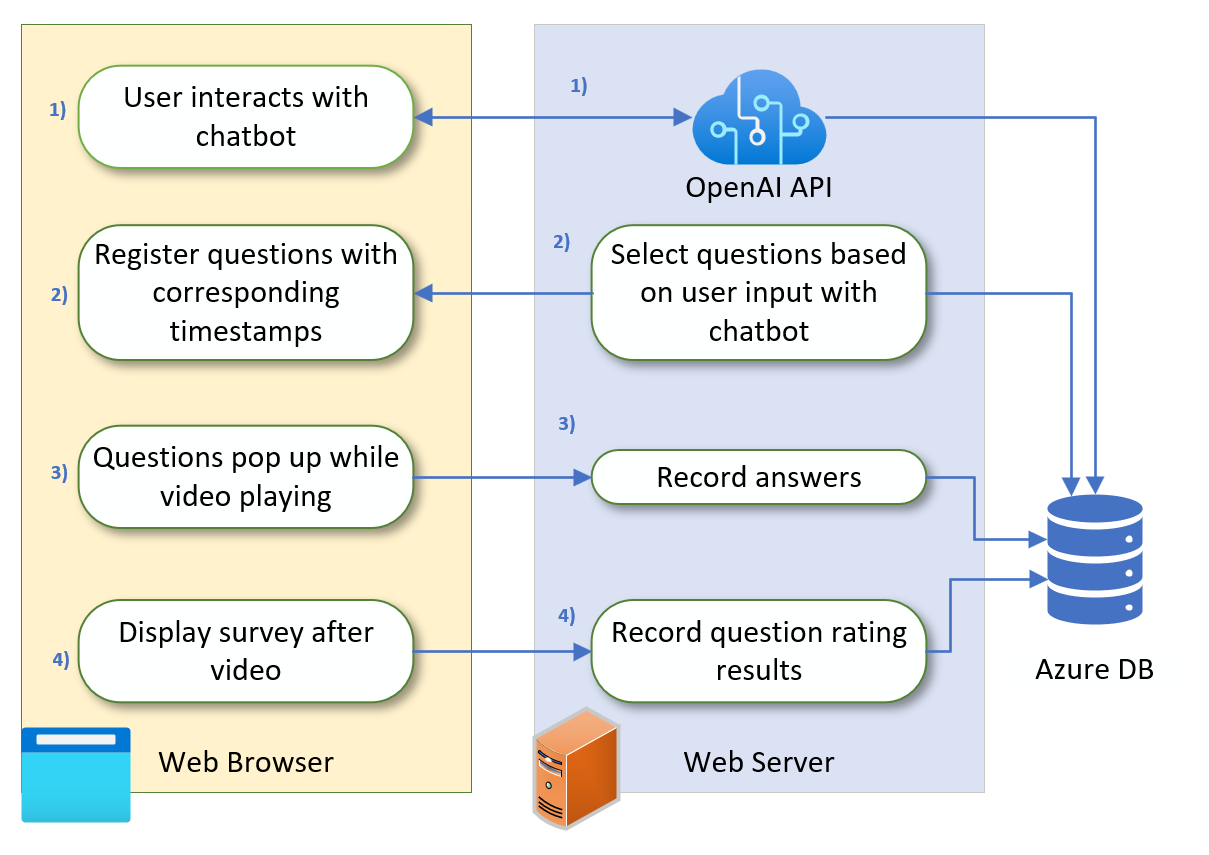} 
    \caption{The architecture flow diagram of the Flask web application illustrates the prototype components, including the web browser, web server, and Azure DB. It highlights four core functionalities: first, participants interact with the OpenAI API chatbot (1); second, the server selects questions based on participant input via the chatbot, and the browser registers these questions with corresponding timestamps (2); third, the browser displays the video and presents questions as the video plays (3); and fourth, the browser presents a question rating page after the video (4). All interaction data and participant choices are recorded in the Azure database.} 
    \Description{}
    \label{fig:system_architecture}
\end{figure}

\end{document}